\documentclass[12pt]{article}

    \usepackage{amsmath,amsxtra,amssymb, amsfonts }

\usepackage{graphicx}

\newtheorem{thm}{Theorem}

 \newtheorem{qst}[thm]{Question}
\newtheorem{lemma}[thm]{Lemma}
\newtheorem{conj}[thm]{Conjecture}

\def\be{\begin{eqnarray}}
\def\ee{\end{eqnarray}}
\def\bee{\begin{eqnarray*}}
\def\eee{\end{eqnarray*}}
\def\bal{\begin{align}}
\def\enal{\end{align}}
\def\pmx{\begin{pmatrix}}
\def\emx{\end{pmatrix}}
\def\bsq{\begin{subequations}}
\def\esq{\end{subequations}}

\def\ds{\displaystyle}
\def\ts{\textstyle}
\def\nn{\nonumber}
\def\vt{\vartheta}
\def\cN{{\cal N}}

\def\wh{\widehat}
\def\raw{\rightarrow}

\def\nrm{\|}
\def\ot{\otimes}

\def\tr{{\rm Tr} \, }

\def\bra{\langle}
\def\ket{\rangle}
\def\kb{ \ket \bra }
\def\wtd{\widetilde}
\newcommand{\proj}[1]{ | #1 \kb  #1|}

\newcommand{\norm}[1]{ \| #1  \|}

\def\mm{ \! - \!}
\def\pp{ \! + \!}

\def\half{{ \tfrac{1}{2} }}

\def\1rt2{\ts \frac{1}{\sqrt{2}} }
\def\rt3{\ts \frac{1}{\sqrt{3}} }

\def\ovb{\overline}

\def\id{{\cal I}}
\def\rmi{{\rm I}}
\def\rmx{{\rm X}}
\def\QC{{\rm QC}}
\def\PD{{\rm PD}}

\def\xeb{{\rm XEB}}
\def\dep{{\rm dep}}

\def\qed{\qquad{\bf QED}}

\def\prf{\noindent {\bf Proof: }}

   \newcommand{\pf}[1]{\noindent {\bf Proof of   #1: }}

 \def\d2{{d ^{\prime}}}
\def\td{\tfrac{1}{d}}

\def\tf4{\tfrac{1}{4}}
\def\tf3{\tfrac{1}{3}}
\def\tf6{\tfrac{1}{6}}

\title{  {\large \bf Pauli Diagonal Channels Constant on Axes}}

\author{ Michael Nathanson \\ {\small
Department of Mathematics and Computer Science} \\
{\small St. Mary's College of California,
Moraga, CA 94575} \\{\small Michael.Nathanson6@stmarys-ca.edu}
  \and Mary Beth Ruskai\thanks{Partially supported  by
    by the National Science  Foundation under Grants  DMS-0314228 and
  DMS-0604900
and by  the National Security Agency  and
Advanced Research and Development Activity   under
Army Research Office   contract number    DAAD19-02-1-0065.}        \\  {\small Department of Mathematics,
     Tufts University,
       Medford, MA 02155} \\
     {\small     Marybeth.Ruskai@tufts.edu}}

\begin{document} 

      \maketitle

      \begin{abstract}
       We define and study the properties of channels which are
       analogous to unital qubit channels in several ways.   A full
        treatment can be given only when the dimension
       $d = p^m$ a prime power,  in which case each of the
       $d+1$ mutually unbiased bases (MUB) defines an axis.    Along  
each
       axis the channel looks like a depolarizing channel, but the
       degree of depolarization depends on the axis.
        When $d$ is not a prime power,  some of our  results still 
hold,
       particularly in the case of channels with one symmetry axis.
       We describe the convex structure of this class of channels
       and the subclass of entanglement breaking channels.  We
       find new bound entangled states for $d = 3$.

       For these channels, we show that the multiplicativity  
conjecture for
       maximal output $p$-norm holds for $p = 2$.  
        We also find channels with behavior not exhibited by unital  
qubit
        channels, including  two pairs of orthogonal bases with
        equal output entropy in the absence of symmetry.   This 
provides
        new numerical evidence for the additivity of minimal output  
entropy.

        \end{abstract}

       \pagebreak
       
       \tableofcontents
       
        \addtolength{\parskip}{+0.9ex}

\section{Introduction}

The results presented here are motivated by the desire to find channels
for dimension $d > 2$ whose properties are similar to those of the unital 
qubit channels, particularly with respect to optimal output purity.   
 A channel is described by a completely positive, trace-preserving 
 (CPT) map.    The channels we construct are similar to unital qubit
channels in the sense that their effect on a density
matrix can be defined in terms of multipliers of components along
different ``axes'' defined in terms of mutually unbiased
bases (MUB).     When all multipliers are positive, these channels are
very much like unital qubit channels with positive multipliers.   However,
when some of the multipliers are negative the new channels can 
exhibit behavior not encountered for unital qubit channels.

For a fixed orthonormal basis ${\cal B} = \{ | \psi_k \ket \}$, the quantum-classical
(QC) channel 
\be
   \Psi^{\rm QC}(\rho) = \sum_k     \bra \psi_k , \rho \, \psi_k \ket  \, \proj{\psi_k} 
\ee
projects a density matrix $\rho$ onto the corresponding diagonal matrix in this
basis.    A convex combination   $\sum_J  t_J \Psi^{\rm QC}_J (\rho)$
of QC channels in a collection of orthonormal bases   ${\cal B}_J = \{ | \psi_k^J \ket \}$
is also a channel; in fact, it is an entanglement breaking (EB) channel.  
We consider channels which are
 a linear combination of the identity map $\id(\rho) = \rho$ and a convex
 combination of QC channels whose bases are mutually unbiased, i.e.,
 satisfy
 \be \label{mubdef}
        |\bra  \psi_m^J,  \psi_n^K \ket |^2 =  \begin{cases}   
               \td  &  \hbox{for}  ~ J \neq K  \\
               \delta_{mn}  &  \hbox{for}  ~ J = K   \end{cases}
 \ee
Such channels can be written in the form
\be   \label{axcdef}
\Phi = s\id + \sum_L t_L  \Psi^{\rm QC}_L
\ee
with
\be   \label{stcond}
s +  \sum_L t_L = 1  \qquad \text{and}   \qquad  t_L \geq 0, \quad s \geq \tfrac{-1}{d -1} .  
\ee
The first condition ensures that  $\Phi$ is trace-preserving  (TP), and the pair that it is completely positive (CP), as will be shown in Section~\ref{sect:2}.

It is well-known that ${\bf C}^d$ can have at most $d+1$ MUB and that
this is always possible when $d = p^m$ is a prime power.  
We are primarily interested in channels of the form \eqref{axcdef} when
such a full set of $d + 1$ MUB exist.    In that case, it is natural to 
generalize the Bloch sphere representation so that a density matrix $\rho$
is represented by a vector $v_{Jj}$ as in  \eqref{rho} and regard each of the 
MUB as defining an ``axis''.   The effect of the channel  \eqref{axcdef}
on a density matrix is to take $v_{Jj} \mapsto    (s + t_J) v_{Jj} $, i.e,
to multiply each $v_{Jj}$ by the number $  \lambda_J = s + t_J$.
Since this action depends only on the axis label $J$ we call these
channels ``constant on axes''.

 In Section~\ref{sect:2} we introduce the relevant notation
 and describe several equivalent ways of representing channels constant
 on axes.     We also describe important subclasses of these channels
 in Section~\ref{sect:sub}  and discuss their structure as a convex
 set in Section~\ref{sect:conv}.     More details about our approach to MUB
 and relevant ways of representing states and channels are given in
 Appendix~\ref{app:pre}.

    In Section~\ref{sect:EB} we study the
entanglement-breaking (EB) subclass, emphasizing conditions on the 
multipliers.   We also give some conditions under which the channels
define bound entangled states when $d = 3$.

In Section~\ref{sect:FH} we study channels which are linear combinations
of the depolarizing channel, the projection onto the diagonal of a matrix
and the completely noisy channel.  These channels have one symmetry axis.
They do not require MUB  for their definition; however, when one has a full set
 of MUB  they can
be rewritten as channels constant on axes.   We give
 necessary and sufficient conditions for the channels to be EB and consider
 their optimal output purity.
 
In Section~\ref{sect:mult} we consider the maximal output purity of channels
constant on axes, particularly the additivity conjecture for minimal entropy
and the multiplicativity of the maximal $p$-norm.    
 We show that for those with some negative multipliers,
the optimal output purity need not occur on the ``longest'' axis.     Indeed,
one can even have two axes with different multipliers for which the 
corresponding MUB have equal output entropy.    Numerical study of such
channels gives new evidence for additivity of the minimal output entropy.
In Section~\ref{sect:posmult} we conjecture that channels with non-negative multipliers achieve their maximal output purity on axis states and explore the connection to multiplicativity. In Section~\ref{sect:p2mult} we show that  multiplicativity  holds for $p = 2$
 for all channels constant on axes and extend this to channels
 constant on the ``longest'' axis.     
 
  The paper contains a number of Appendices, the first of which is primarily
  expository.   The first two sections of 
  Appendix~\ref{app:pre} describe representations of states and channels
  from the perspective that the $d \times d$ matrices form a Hilbert space
  with the inner product $\bra A , B\ket = \tr A^\dag B$.   Section~\ref{sect:paul}
discusses expansions in generalized Pauli matrices and their connection to MUB.
Section~\ref{sect:mub}  gives more information about MUB;
Section \ref{app:MUBchan} considers some alternative ways
of using MUB  to describe channels; and   
Section~\ref{app:sing_conj} considers channels which are formed
from conjugations on a single axis.    Finally, a simple proof of the so-called
computable cross norm (CCN) condition is given in Section~\ref{app:ccn}.

  The remaining appendices contain details of proofs which are omitted in the main text.
  Appendix~\ref{app:mult} contains several proofs related to the multiplicativity conjecture.
  Appendix~\ref{app:FHsep} proves separability of certain state representatives 
  which determine the EB region for channels with one symmetry axis.
  Appendix~\ref{app:CJ} describes the state representative when $d$ is prime.
  Appendices~\ref{sect:d3} and \ref{app:dprime} use this result to
  obtain extreme points of the EB region for prime $d$ as well as the PPT region for the case $d=3$. 

\section{Channels constant on axes} \label{sect:2}

\subsection{Notation and generators of MUB} \label{sect:W}

For any collection ${\cal B}_J = \{ | \psi_k^J \ket \}$ of orthonormal bases on
${\bf C}^d$, we can define the operators
\be   \label{Wgen}
     W_J = \sum_{k=1}^d \omega^k \, \proj{ \psi_k^J} , \qquad   J = 1,2, \ldots d \pp 1
\ee
where $\omega = e^{2 \pi i /d}$.    It follows that  
  \be   \label{ax.state}
    \proj{\psi_n^J} = \td   \sum_{j = 0}^{d-1}  \ovb{ \omega}^{nj} \, W_J^j 
     = .\td \Big[ I +   \sum_{j = 1}^{d-1}  \ovb{ \omega}^{nj} \, W_J^j \Big].
       \ee
By construction, $\tr W_J^m = 0 $ for $m =1,2, \ldots d \mm 1$,  
$W_J^m$ is unitary for any integer $m$, $(W_J^m)^\dag = W^{-m} = W^{d-m}$,
and each of the  operators  $W_J$  generates  
a cyclic group  of order $d$.  If, in addition, the bases are mutually
unbiased \eqref{mubdef} then when $J \neq K$ and $m$ and $n$ are not both zero 
\be   \label{Worth}
    \tr (W_J^{m})^\dag W_K^n = \sum_{jk} \omega^{jm-kn} |\bra \psi_j^J, \psi_k^K \ket|^2 =
     \td \sum_{j = 1}^d   \omega^{jm} \sum_{k = 1}^d  \ovb{\omega}^{kn} = 0.
\ee
When there are $d + 1$ MUB, this gives $d^2 - 1$ unitary operators
$\{ W_J^m \}_{m = 1 \ldots d \mm 1, J = 1\ldots d+1}$ which satisfy the orthogonality
condition    $ \tr W_J^{d-m} W_K^n = d \, \delta_{JK} \delta_{mn}$ and, hence, form an 
orthogonal basis for the subspace of trace zero matrices in $M_d$.   
Note that this immediately
implies that $d + 1$ is the maximum number of MUB for ${\bf C}^d$.     

We call the unitary operators $W_J$ the {\em generators} of the MUB.   
When we   have a full set of $d+1$ MUB, adding the identity
matrix $\rmi$ to $\{ W_J^m \} $ gives an orthogonal basis of unitaries (OBU) for 
$M_d$ with norm $d^{1/2}$.    Therefore, any density matrix $\rho$ can be written as
\be   \label{rho}
    \rho ~ = ~    \td \Big[ I + \sum_{J = 1}^{d + 1} \sum_{j = 1}^{d-1}  v_{J j} W_J^j \Big]   
    \ee 
    with  $v_{J j}  = \tr W_J^{-j} \rho $.  This is  the standard expansion
    of a vector in a Hilbert space using an orthogonal basis; the only novelty is that our
   Hilbert space is the set of $d \times d$ matrices $M_d$ with the Hilbert-Schmidt
   inner product $\bra A , B\ket = \tr A^\dag B$.   Equation \eqref{rho}
   can also be considered a generalization
   of the Bloch sphere representation.  Both  viewpoints are considered in more detail
   in   Appendix~\ref{app:pre}.
 It is straightforward to show that  
  \be   \label{QCdef}
 \Psi_K^{\rm QC}(\rho) = \td \sum_{j=0}^{d-1} W_K^j \rho W_K^{-j} = 
      \td \Big[ I +  \sum_{j = 1}^{d-1}  v_{K j} W_K^j \Big]   .
 \ee
 This says that the effect of   $\Psi_K^{\rm QC}(\rho) $
is simply to multiply $v_{Jj}$ by $1$ for $J = K$ and by $0$ for $J \ne K$. 
Since \eqref{QCdef} has the Kraus operator sum form, the Kraus operators for
a QC channel corresponding to the basis  ${\cal B}_J $  can be chosen as  
$\tfrac{1}{\sqrt{d}} W_J^j, j = 1,2, \ldots d$.

 \subsection{Equivalent representations}
 
 The results of the previous section allow us to give some
equivalent ways of writing channels  constant on axes.   First, 
observe that a map of the form \eqref{axcdef}
   can  be written as 
   \be    \label{map:conjW}
    \Phi(\rho)   = a_{00}\, \rho +  \tfrac{1}{d-1} \ds{ \sum_{J = 1}^{d + 1} \sum_{j = 1}^{d-1} }
                              a_J \, W_J^j \rho W_J^{-j} .
   \ee 
with  $ a_{00} = s + \td \sum_J t_J = \td\big[(d \mm 1) s + 1 \big]$
and  $a_J  = \tfrac{d-1}{d} t_J $.   In this form the TP condition in \eqref{stcond}
becomes  $a_{00} +  \sum_J a_J = 1$ and
the next pair of conditions are equivalent to $a_{00} \geq 0$ and
$a_J \geq 0$ for all $J$.   Then \eqref{map:conjW} has the  operator sum form
of a CP map with Kraus operators $\sqrt{a_{00}} I $ and $\sqrt{ a_J/(d -1)} \, W_J^j$.   
Thus, the conditions \eqref{stcond} suffice for $\Phi$ to be CPT.   
 It follows from Theorem~\ref{thm:aCP} in Appendix \ref{sect:OBU2} that the converse is also true, i.e., a map
 of the form \eqref{map:conjW} is not CP unless $a_{00} \geq$ and 
$a_J \geq 0$ for all $J$.
  
 It follows from the comment after \eqref{QCdef}  
 that the effect of  a map of the form \eqref{axcdef} 
 can be expressed as
\be   \label{multform} 
     \Phi:    \td  \Big[ I + \sum_{J = 1}^{d + 1} \sum_{j = 1}^{d-1}  v_{J j} W_J^j \Big]   ~
       \longmapsto ~ \td \Big[ I + \sum_{J = 1}^{d + 1} (s + t_J) \sum_{j = 1}^{d-1}  v_{J j} W_J^j \Big]   .
\ee
so that $v_{Jj} \mapsto  \lambda_J  \, v_{Jj} $ with $\lambda_J = s + t_J$.
Thus,  every such channel corresponds to a unique vector  in ${\bf R}^{d+1}$
which we write as $[\lambda_1, \lambda_2, \ldots \lambda_{d+1}]$  with
$\lambda_J = s + t_J$.     When all of the $\lambda_J$ are equal, the channel is
depolarizing.    Thus, another view of a channel constant on axes is that  
an input    on the $J$-th axis has   the same ouput as a depolarizing channel 
  with   $\lambda = \lambda_J$ in   \eqref{sub:dep}.   This follows immediately
  from \eqref{multform} and the fact that $\gamma = \sum_n \mu_n \proj{\psi_n^J}$ 
    has $v_{Lj} = 0$ for $L \neq J$.

\begin{thm}
Let  ${\bf C}^d$ have a full set of $d \pp 1$ MUB and let
$[\lambda_1, \lambda_2, \ldots, \lambda_{d+1}]$ be a vector in ${\bf R}^{d+1}$.   
Then \eqref{multform} defines a CPT map  if and only if
\be  \label{lamineq}
 - \, \tfrac{1}{d-1} ~ \leq ~  \sum_{J} \lambda_J  ~ \leq ~   1 + d \, \min_K \lambda_K  . 
\ee
\end{thm}
\prf If one uses the TP condition in \eqref{stcond} to eliminate
$s$, the two CP inequalities are equivalent to
\bsq   \label{lamineqq}   \be    \label{lamineqa}
        \sum_{J \neq K} \lambda_J  & \leq & 1 + (d-1) \lambda_K   \qquad K = 1,2,\ldots d \pp 1  \\
        \sum_J  \lambda_J  & \geq & - \tfrac{1}{d-1}.   \label{lamineqb}
\ee \esq
which is clearly equivalent to \eqref{lamineq}.
\qed

\subsection{Subclasses}   \label{sect:sub}

We now describe some important subclasses of   channels constant on axes:

\begin{enumerate}
    \renewcommand{\labelenumi}{\theenumi}
    \renewcommand{\theenumi}{(\alph{enumi})}
 
\item {\bf QC channels:}  Let $\Psi_L^{\rm QC}$ have the form \eqref{QCdef}.
Then its   multiplier is   $[0, \ldots, 1, \ldots, 0]$
and  $ a_{00} = a_L = \td$ and $a_{K j} = 0$ for $K \neq L$.   
  
 \item {\bf Phase damping channels:}  Let 
 $\Psi_{L,\lambda}^{\rm PD} = \lambda \id + (1-\lambda) \Psi_L^{\rm QC}$ with
 $ -\tfrac{1}{d-1} \leq \lambda \leq 1$.   Then  $\Phi$ has a multiplier of the form 
   $[\lambda, \ldots \lambda, 1, \lambda, \ldots, \lambda]$
  and 
  $ a_{00} = \lambda  + \tfrac{1 - \lambda}{d}$, $a_{L j} = \ \tfrac{1 - \lambda}{d},$ and 
  $a_{K j} = 0$ for $K \neq L$.    
    The $d$ axis states which are eigenvectors of $W_L$ are invariant
  and thus have pure outputs.

  \item {\bf Extreme phase damping channels:}    Let $\Psi_L^{\rmx} = \Psi_{L,\lambda}^{\rm PD}$
   with $\lambda =   \tfrac{-1}{d-1} $ so that  $a_{00} = 0$ and
    $a_{Kj} =   \tfrac{-1}{d-1}  \delta_{KL}$.    Since no axis channel (except $\id$)
    can have fewer non-zero $a_{Kj}$, these channels   are extreme points of the 
             convex set of  axis channels.    
    Each  $\Psi_L^{\rmx}$  has a multiplier has the form
   $[ \tfrac{-1}{d-1}, \ldots, \tfrac{-1}{d-1}, 1, \tfrac{-1}{d-1}, \ldots \tfrac{-1}{d-1}]$
   with $1$ in the $L$-th position.
 When $d = 2$, each $\Psi_L^{\rmx} $ is a conjugation with one of the Pauli
  matrices $\sigma_L$ and its multiplier is a permuation of $[-1,1,-1]$.
  
        \item  {\bf Extreme EB channels:}   The channels
        $\Psi_L^\xeb \equiv \tfrac{-1}{d - 1} \Psi_L^\QC +  \tfrac{1}{d-1} \cN = 
   \td  \sum_{J \neq L}  \Psi_J^{\rmx} $Ê  have multiplier 
   $[0, \ldots 0, - \tfrac{1}{d-1}, 0 \ldots 0]$ and are extreme points
   of the set of EB channels.    The  channels we denote
   $\Psi_L^{\rm YEB}$ have multipliers which are permutations of 
$  \big[\tfrac{d-2}{2(d-1)}, \tfrac{-1}{2(d-1)} ,\ldots  \tfrac{-1}{2(d-1)} \big]$;
   for $d > 2$ these are also extreme points of the set of EB
   channels, as will be shown in Section \ref{sect:EB}.

\item {\bf Depolarizing channels:} The channel
$     \Psi^{\rm dep}_\lambda(\rho) = \lambda \rho + (1 - \lambda) \td \rmi $
has multiplier   $[\lambda, \lambda, \ldots, \lambda]$ and can be written as 
\be  \label{sub:dep}
 \Psi^{\rm dep}_\lambda \, = \,  \lambda \id + (1 - \lambda) \cN ~ = ~ 
  \sum_{L=1}^{d+1}  \tfrac{1}{d+1}\,  \Psi_{L,\zeta}^{\rm PD}
  \ee 
  with $\zeta = \frac{ \lambda(d+1) - 1}{d}$.  Then  $a_{00} = \lambda + \tfrac{1 - \lambda}{d^2}$
   and $a_{L j} =  \tfrac{1 - \lambda}{d^2}$.
   
   \item {\bf Channels with one symmetry axis:} 
   Channels of the form \eqref{axcdef}
with all but one of the $t_J$ identical have multipliers
    $[\lambda , \lambda , \ldots ,\lambda ,\eta, \lambda, \ldots \lambda]$ 
    with   $\eta =  s + t_L$.   They are naturally 
     regarded as ``squashed'' when $0 < \eta < \lambda$.     In general,
     they are symmetric with respect to ``rotations'' about the   special
     axis $L$.
      The boundary case with one $t_L = 0$ has $\eta =  \tfrac{d \lambda - 1}{d -1 }$. 
  This   is called a   ``two-Pauli'' channel in the qubit case; we call them
  ``maximally squashed''.   
       These channels can be written in several equivalent forms
\be  \label{AD}  
      \Psi_{L,\lambda}^{\rm MxSq}  & =   & \sum_{K \neq L }   \td  \, \Psi_{K,\zeta}^{\rm PD}   
            ~   = ~ \zeta \id +  \sum_{K \neq L }   \td  \, \Psi_{K,\zeta}^{\rm QC}    
            ~   = ~  \lambda \id + \tfrac{1- \lambda}{d} \sum_{J \neq L} \Psi_L^\rmx   \\ \nn
         & = &  \lambda \id + (1 - \zeta) \cN -
          \tfrac{1-\zeta}{d} \Psi_L^{\rm QC}   
          ~   = ~              \Psi^{\rm dep}_\lambda  
            +  \tfrac{1-\zeta}{d}  \big(\cN -  \Psi_{L }^{\rm QC} \big) 
              \ee 
with $1 \geq \lambda = \zeta + \td (1 - \zeta) \geq 0$.  

 \item   For qubits, the channel  which takes
   \be
        \rho \mapsto  \sigma_J \Psi^{\rm dep}_\lambda \sigma_J = \lambda
            \sigma_J \rho \sigma_J +(1 - \lambda) \half \rmi  \qquad  J = 1,2,3
   \ee
   can be thought of as depolarizing from conjugation with $\sigma_J$; its 
    multiplier has the form
      $[-\lambda, +\lambda, -\lambda]$ (with the $+$ sign in the $J$-th position).
   For $d > 2$, this has no direct generalization,  but one might consider
   channels which ``depolarize''  from the other extreme points, e.g., 
   $\lambda   \Psi_L^{\rmx} + (1 - \lambda) \cN $
    which has  multiplier $\big[\lambda,\tfrac{-\lambda}{d-1}, \ldots   ,\tfrac{-\lambda}{d-1} \big] $.
These channels are also a subclass of those with one symmetry axis. 
 
    \end{enumerate}
    
     \begin{table}[here]
$$  \begin{array}{|cc|cc|cc|}  \hline
&\hbox{description}  & & \hbox{qubit} &  & d > 2   \\[0.05cm]  \hline   
&  \hbox{Identity} & \id & [1,1,1] & \id & [1,1, \ldots 1]    \\[0.3cm]   
&  \hbox{ max noise} & \cN & [0,0,0] & \cN & [0,0, \ldots 0]    \\[0.3cm]   
     \hbox{(a)}  &   QC   & QC & [1,0,0] &  \Psi_L^{\rm QC} &  \big[ 1,0, \ldots 0\big]     \\[0.3cm]  
     \hbox{(b)}   &  \hbox{phase-damping} &   & [1, \lambda, \lambda]   
   & \Psi_{L,\lambda}^\PD & [1, \lambda, \ldots \lambda]    \\[0.3cm]  
   \hbox{(c)} &  \hbox{extreme points} & \sigma_L \rho \sigma_L & [1,-1,-1] 
   & \Psi_L^\rmx & \big[1,\tfrac{-1}{d-1}, \ldots   ,\tfrac{-1}{d-1} \big]  \\[0.3cm]  
    \hbox{(d)}  &  \hbox{extreme EB}   &   \sigma_j  QC \sigma_j  & [-1,0,0] &  \Psi_L^\xeb &  \big[ \tfrac{-1}{d-1},0, \ldots 0\big]     \\[0.3cm]  
         &    \hbox{extreme EB for } d>2   &     & [-\tfrac{1}{3},\tfrac{1}{3} ,\tfrac{1}{3}] &  \Psi_L^{\rm YEB}&  
    \big[\tfrac{d-2}{2(d-1)}, \tfrac{-1}{2(d-1)} ,\ldots  \tfrac{-1}{2(d-1)} \big]
   \\[0.3cm]  
    \hbox{(e)} &  \hbox{depolarize from $\id$} & \Psi^\dep_{\lambda} & [\lambda, \lambda, \lambda]  
     & \Psi^\dep_{\lambda} & [\lambda, \lambda, \ldots \lambda]  \\[0.3cm]   
\hbox{(f)}   &  \hbox{max squashed} &  \hbox{two-Pauli}  &  [2 \lambda \mm 1, \lambda, \lambda] 
   & \Psi_{L,\lambda}^{\rm MxSq} & \big[ \tfrac{d \lambda - 1}{d-1}, \lambda,  \ldots \lambda \big]
 \\[0.3cm]     \hbox{(g)}  &   \hbox{depolarize from $\Psi_L^\rmx$} & & [\lambda,- \lambda, -\lambda]  
     &   &
     \big[\lambda,\tfrac{-\lambda}{d-1}, \ldots   ,\tfrac{-\lambda}{d-1} \big]   \\[0.3cm]     \hline   \end{array}$$  \vskip-0.1cm
\caption{Comparison of axis channels classes  to unital qubit classes} \label{tab:compar}
\end{table}

\subsection{Convex structure}  \label{sect:conv}
 
 The set of axis channels is convex, and we have already noted
 that the extreme phase-damping channels 
 $\Psi_{L}^\rmx = \tfrac{-1}{d-1} \id + \tfrac{d}{d+1} \Psi_L^\QC$ are
 extreme points of this set.   Adding the identity $\id$ gives
 all the extreme points.
 \begin{thm}  \label{thm:ext}
When ${\bf C}^d$ has $d \pp 1$ MUB, the set of channels constant on axes 
is the convex hull of the identity
 $\id$ and the extreme phase-damping channels  $\Psi_{L}^\rmx$.
 \end{thm}
\prf   It suffices to observe that when $\Phi$ is given by \eqref{axcdef},
it  can be written as
\be   \label{conv}
 \Phi = a_{00} \id + \sum_{J = 1}^{d+1}    a_J \Psi_{J}^\rmx  ~ = ~
    \td[(d \mm 1) s + 1] \id + \tfrac{d -1}{d}   \sum_{J = 1}^{d+1}   t_J  \Psi_{J}^\rmx
 \ee
with coefficients as in \eqref{map:conjW} and \eqref{axcdef} respectively.   
The TP and CP conditions \eqref{stcond} imply that the coefficients sum to $1$
and are nonnegative.  \qed

Each of the $d+2$ inequalities \eqref{lamineqq}  defines a half space
corresponding to the hyperplane defined by $d +1$ of the extreme points
in Theorem \ref{thm:ext}.   Then the intersection of these half-spaces yields
the convex set of channels constant on axes. 
When $d = 2$, \eqref{lamineqq} is equivalent to
$|\lambda_j \pm \lambda_k| \leq |1 \pm \lambda_{\ell}|$ for $j,k,\ell$ distinct;
which are the conditions  \cite{FA,KR1,RSW} needed to
 ensure that a unital qubit channel is CP. 

It is now well-known \cite{FA,KR1,RSW}
that the multipliers $[\lambda_1,\lambda_2, \lambda_3]$
 for the unital qubit channels form a tetrahedron  with vertices at
 $[1,1,1], [1,-1,-1], [-1,1,-1]$, and $[-1,-1,1]$, and that the subset of 
 entanglement breaking (EB)  channels corresponds to the octahedron
 obtained from the intersection of this tetrahedron with its
 inversion through the origin.    Removing this octahedron
 leaves 4 disjoint sets (also tetrahedrons) which can be
 transformed into one another by conjugation with the Pauli
 matrices $\sigma_J$.    Each of these sets has multipliers with
 fixed signs determined by one of the maps  $\Psi_J^\rmx$
 and is the convex hull of this map and three QC channels.
 For example,   the set with only $\lambda_1 \geq 0$ is the
 convex hull of  $\Psi_1^\rmx(\rho) = \sigma_1 \rho \sigma_1$
 with multiplier $[1,-1,-1]$  and the QC maps with multipliers
 $[1,0,0], [0,-1,0], [0,0,-1]$.
 
  \begin{figure}[here]
 \begin{center}
    \includegraphics*[width=4cm,height=4cm,keepaspectratio=true]{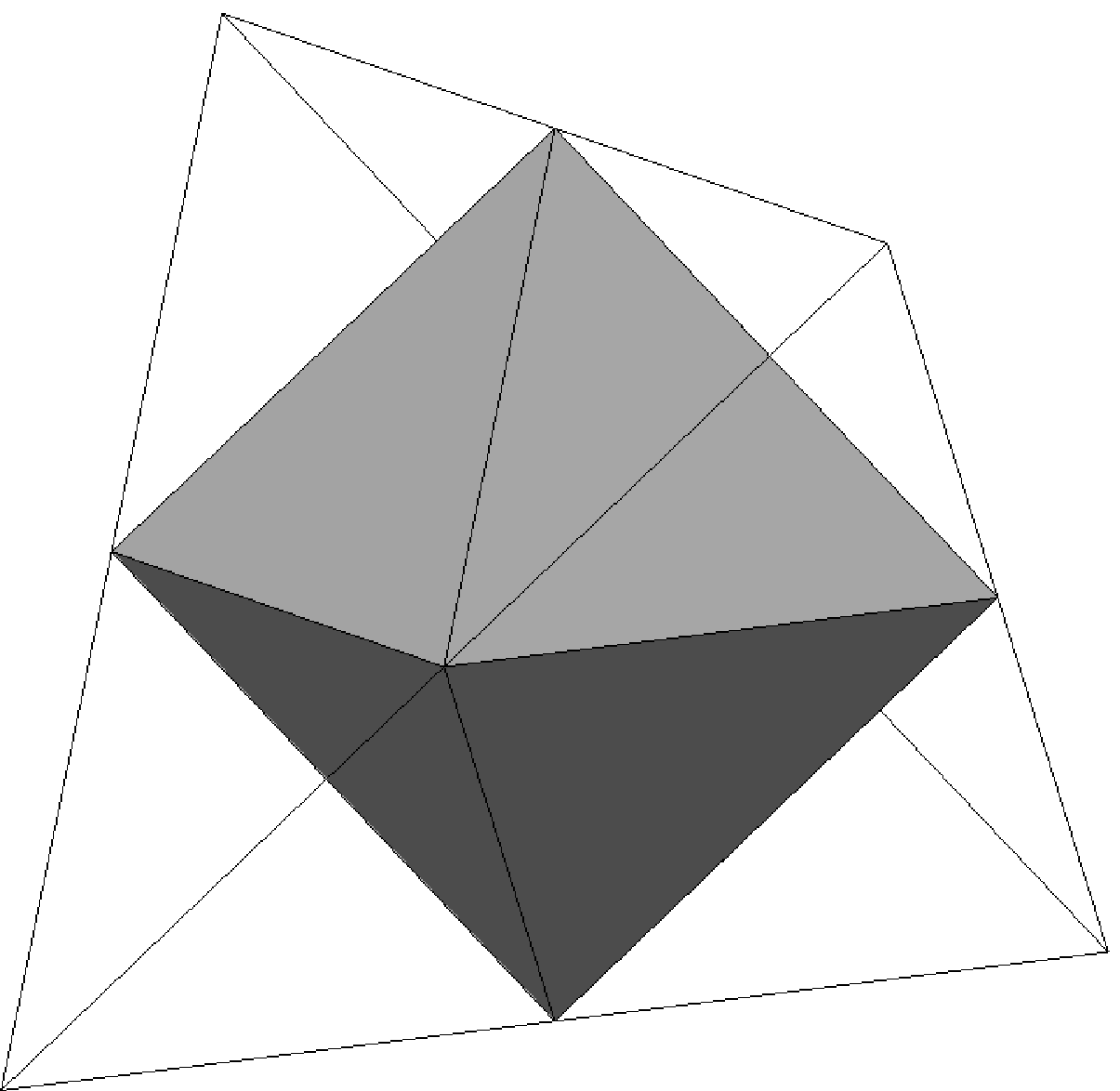} 
\end{center}
 \caption{The tetrahedron of qubit channels and the octahedral EB subset} \label{fig:qub}
 \end{figure}
 
 When $d > 2$, this picture is modified significantly. The set of CPT maps is still
  the convex hull of  $\id$ with multiplier $[1,1, \ldots, 1]$
 and  $d+1$ channels whose multipliers are permutations of
 $ \big[1,\tfrac{-1}{d-1}, \ldots   ,\tfrac{-1}{d-1} \big]$.
 However, the identity is a distinguished vertex from which  edges extend to give
 the $d+1$ lines of phase-damping channels between $\id$ and $\Psi_L^\rmx$.
 One also has a ``base'' formed from  the edges connecting pairs of the latter.    There
 is only one symmetry axis.   After
 removing the EB channels,  one again has a disjoint region $\Delta_0$ 
 which contains the non-EB channels with all  multipliers non-negative;
 this is the convex hull of $\id$ and the $\Psi_L^{\rm QC}$, as before.    However, the
 picture with negative multipliers is far more complex.
  
 For $d = 3$, the ``base'' corresponds to the tetrahedron given by the
 convex hull of the four vertices  $\Psi_L^{\rmx} $ with $L = 1,2,3,4$.
 The center of each of the four faces is  
 $\Psi^{\xeb}_L  = \td  \sum_{J \neq L}  \Psi_J^{\rmx} $.   Since
 this is EB, the tetrahedron
 obtained by joining these four points (which is the set of channels with all
 $\lambda_k \leq 0$)  is a subset of the EB channels.
 However,  it follows from Theorem~\ref{thm:EBT<1} in Section \ref{sect:conv} that no point on an 
 edge connecting two $\Psi_L^{\rmx}$ is EB which means that, 
unlike the qubit case, removing the EB channels
 from the base does not leave $d$ disjoint sets.    This argument
 extends to all $d > 2$.
 
 \begin{figure}[here]  
 \begin{center}
\includegraphics*[width=4cm,height=4cm,keepaspectratio=true]{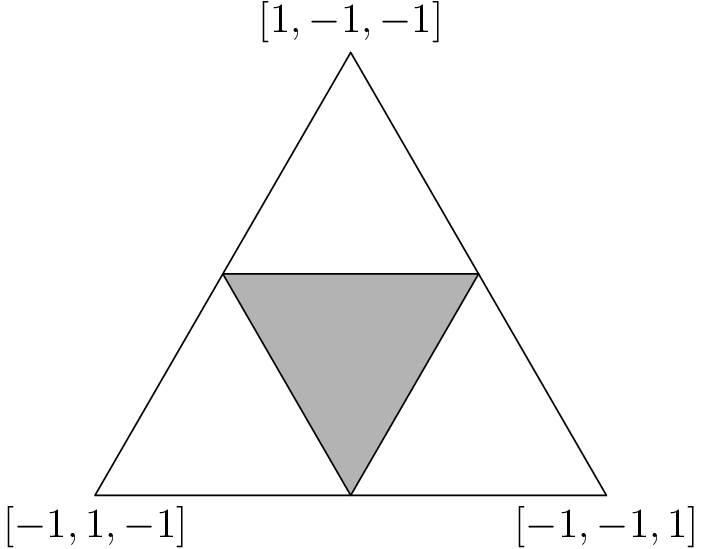} \qquad     \includegraphics*[width=4cm,height=4cm,keepaspectratio=true]{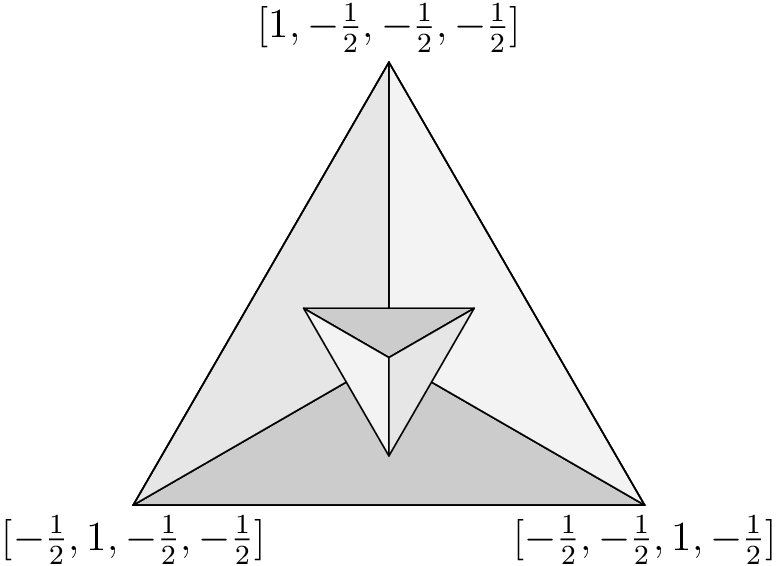} \qquad  \includegraphics*[width=4cm,height=4cm,keepaspectratio=true]{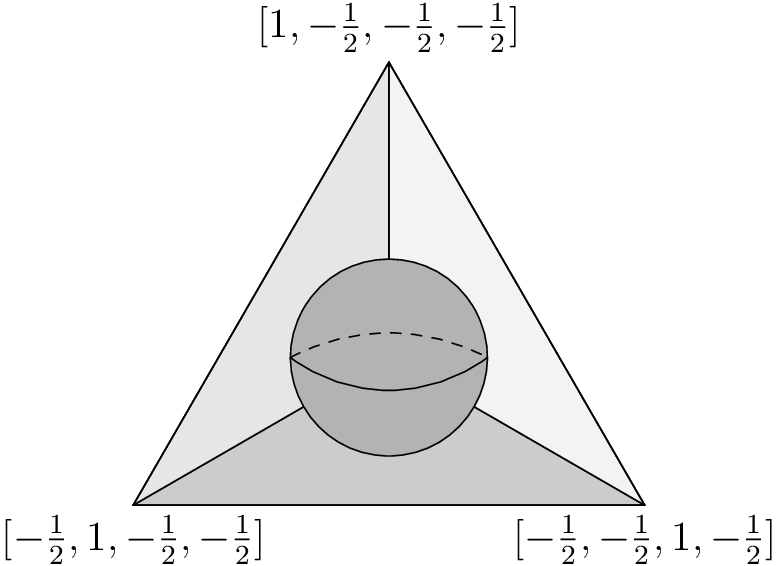}

 \hskip 0.7cm $d = 2$  \hskip 3.4 cm   $d = 3$: EB \hskip 3cm   $d = 3$: PPT 
\end{center}
  \caption{Base of qubit and qutrit channels constant on axes showing
  subregions with all $\lambda_J \leq 0$.   For qubits  this is also the EB region.
  For qutrits, the tetrahedron lies strictly within the EB region; the sphere shows the qutrit PPT channels, as discussed in Section \ref{Section:d3PPT}}.  \label{fig:base}
\end{figure} 

\section{Entanglement breaking channels}  \label{sect:EB}

\subsection{General considerations}

A channel is called entanglement breaking (EB) if its action on half of an entangled
state is separable.     In fact, it suffices to consider its action on the maximally
entangled state $|\beta \ket =  \tfrac{1}{\sqrt{d}}\sum_k |e_k \ot e_k \ket$.   It is well-known that
$\Phi \leftrightarrow \td \sum_{jk} |e_j \kb e_k|  \ot \Phi( |e_j \kb e_k| ) =  (\id \ot \Phi)\proj{\beta}$ 
gives a one-to-one correspondence between CPT maps
taking ${\bf C}^d \mapsto {\bf C}^d$ and density matrices on ${\bf C}^{d^2}$.
The latter  is called the Choi-Jamiolkowski (CJ) matrix or state representative
of the channel.  In \cite{HSR}, it was shown that a channel is EB if and only if its 
CJ matrix  is separable.   
 
 A  channel of the form  $\Phi = t_L \Psi_L^\QC +  (1- t_L) \cN$ is clearly EB 
when $t_L \geq 0$.   However, even for the larger CP range 
 $ \tfrac{-1}{d - 1}  \leq t_L \leq 1$ it is EB  because  
  \be   \label{EBbd}
 \Phi(|\psi_j^L \kb \psi_k^L|) = \delta_{jk} \sum_n \big( \td (1-t_L)  + t_L \delta_{jn} \big) \proj{ \psi_n^L}.
 \ee
This implies that the CJ matrix is diagonal in the product basis $\{|\psi_j^L \ot\psi_k^L\ket\}$
and hence separable.  
 The CP  endpoints of this line are $\Psi_L^\QC$  for $t_L = 1$ and
 $\Psi_L^\xeb \equiv \tfrac{-1}{d - 1} \Psi_L^\QC +  \tfrac{1}{d-1} \cN $ for
 $t_L =  \tfrac{-1}{d - 1}$.   
 
    The positive partial transpose  condition (PPT)  condition
for separability applied to the CJ matrix of a channel says that
$d({\cal T} \ot \Phi)\proj{\beta} =  \sum_{jk} |e_k \kb e_j|   \ot \Phi( |e_j \kb e_k| )$
is positive semi-definite.    This is a necessary condition for a channel to be EB.
By applying the PPT condition to the phase-damping
channel $\Psi^\PD_{L,\lambda}$, one can see that it is EB if and only 
 if $\lambda = 0$.

 It seems natural to conjecture that    $\Psi_L^\QC$ 
 and  $\Psi_L^\xeb$ are the only extreme points of the convex set of EB channels
 constant on axes.    However, this is not the case, as one can see from the
 following theorem which is proved in Appendix~\ref{sect:d3}, where we also 
 show that $\Psi_L^\xeb$ is a true extreme point for any prime $d$.
 \begin{thm} \label{thm:extEB}
 When $d = 3$, the channels $\Psi_L^\QC$, $\Psi_L^\xeb$ and $\Psi_L^{\rm YEB}$
  are extreme points of the convex set of EB.   
 \end{thm}
 For $d = 2$, only the first two channels give extreme points; the channel
 $\Psi_L^{\rm YEB}$ is on the boundary of EB subset, but not extreme.   
 For $d = 3$, it seems natural to conjecture
 that set of EB channels constant on axes is the convex hull of the channels
 in Theorem~\ref{thm:extEB}.    However, it appears that there are regions
 of strict convexity for the PPT condition which yield  additional extreme
 points for $d \geq 3$.      
 
The channels  $\Psi_L^{\rm YEB}$  are considered in Section~\ref{sect:FHEB} where they
are shown to be extreme points of the subset of EB channels  with one symmetry axis.
For $d = 3$, the channel  $\Psi_L^{\rm YEB}$  
 has multiplier  $ [ \tfrac{+ 1}{~4}, \tfrac{-1}{~4},  \tfrac{-1}{~4},\tfrac{-1}{~4}]   $.
 For $d = 4$  it has multiplier
   $ [ \tfrac{+ 1}{~3}, \tfrac{-1}{~6},  \tfrac{-1}{~6},\tfrac{-1}{~6},\tfrac{-1}{~6}]   $; the channel
   with multiplier  $ [ \tfrac{+ 1}{~5}, \tfrac{-1}{~5},  \tfrac{-1}{~5},\tfrac{-1}{~5}, \tfrac{-1}{~5}]   $
   is not CP because $\sum_k \lambda_k = - \tfrac{3}{5} < - \tfrac{1}{3}$ violates
   condition \eqref{lamineqb} and the channel with multiplier   
   $ [ \tfrac{+ 2}{~9}, \tfrac{-2}{~9},  \tfrac{-2}{~9},\tfrac{-2}{~9}, \tfrac{-2}{~9}] $ is EB but not extreme.

  \subsection{Multipliers for Entanglement-Breaking maps}
 
 We now consider EB criteria which can be stated in terms of the multiplier
 for a channel constant on axes.     Any EB channel constant on axes must
 have  $\sum_L | \lambda_L| \leq 1$.    
 This is an immediate consequence of the more general requirement that
 $\norm{ \Phi}_1 \leq 1$ for any EB channel.   This is equivalent to what is
 sometimes called the ``computable cross norm'' (CCN) condition or
 ``rearrangement criterion'' for separability.   However, 
as explained in Appendix~\ref{app:ccn},  this condition
 can be applied directly to the channel without computing  its CJ matrix
 or performing any type of rearrangement.   
  
 \begin{thm}  \label{thm:EBT<1}
Let $\Phi$ be  a channel constant on axes.  
  If $\Phi$ is EB, then $\sum_L | \lambda_L| \leq 1$.    Moreover,
  if all multipliers $\lambda_k \geq 0$, then $\Phi$ is EB if and only
  if $\sum_L \lambda_L \leq 1$.
\end{thm}
 
\pf{Theorem \ref{thm:EBT<1}} The necessity follows immediately 
from Theorem~\ref{thm:EBT.pauli}
and the fact that the singular values, $\phi_s$, of $\Phi$ are    $  | \lambda_L |$, each with
degeneracy $d \mm 1$, and $1$ which is non-degenerate.  Thus  
\bee   
d \ge \sum_s  |\phi_s| =   1 + (d-1)\sum_{L} | \lambda_{L} |.
\eee
Sufficiency follows immediately from the fact that when all
$\lambda_L \ge 0$ and $\sum_L \lambda_L \leq 1$,  one can write
\be
   \Phi = \sum_L  \lambda_L \Psi^\QC_L  + \big(1 - \sum_L \lambda_L \big) \cN
\ee
as a convex combination of EB channels.   \qed

 When an EB channel is written   in the form \eqref{axcdef},  we see that $s \leq  0$.   This is
 an immediate corollary of Theorem~\ref{thm:EBT<1} and
   \be   \label{eq22}
        1 \geq  \sum_J |\lambda_J | \geq   \sum_J \lambda_J = (d+1) s + \sum_J t_J = ds + 1.
   \ee
   
In \cite{EBQ} it was shown that a  unital qubit channel is always EB when
some $\lambda_k = 0$.   
It follows from  \eqref{lamineqa} that if the smallest $\lambda_J = 0$
then  $\sum_{J \neq L}  \lambda_J \leq 1$ so that the channel is EB,
giving a partial extension of the qubit result.   However,   
  when some  $\lambda_k $ are negative a channel with some
  $\lambda_J = 0$ need not be EB  as shown by the
following example for $d = 3$:  
\be  \label{EB0}
    \Phi =  \tfrac{2}{3} \Psi_1^{\rm X} +  \tfrac{1}{3} \Psi_2^{\rm X}  =
   [\half,0,-\half,-\half]
\ee
Since $\sum_k |\lambda_k| = \tfrac{3}{2} > 1$, Theorem~\ref{thm:EBT<1} implies 
that this channel is not EB.

When all $\lambda_k \leq 0$, inequality  \eqref{lamineqb} implies that
$\sum_k  |\lambda_k| \leq \tfrac{1}{d-1}$ and hence that $\Phi$ is in the 
convex hull of $\cN$ and the set $\{\Psi_J^\xeb \}$.    Thus every channel with all 
 $\lambda_k \leq 0$ is EB.   What remains is to find precise necessary
 and sufficient conditions for a channel with both positive and negative
 multipliers $\lambda_L$ to be EB. In Figure~\ref{fig:boundent}, there are channels with $\sum_J | \lambda_J | \leq 1$ which lie outside the PPT region; thus we see the condition from Theorem~\ref{thm:EBT<1} is not sufficient for EB. 
 
 Channels with exactly one symmetry    
axis (i.e., those for which $d$ of the $\lambda_J$ are equal) are studied in
Section~\ref{sect:FH}, in which we show that 
  $\sum_L | \lambda_L| \leq 1$ is necessary and sufficient for 
  $\Phi$ to be EB.  
      When $d = 3$, this implies that the
       channel  $\Psi_L^{\rm YEB} = [\tfrac{1}{4},  -\tfrac{1}{4}, -\tfrac{1}{4},-\tfrac{1}{4}]$ is EB.
      This channel is outside the convex hull of $\Psi_L^\QC$ 
 and  $\Psi_L^\xeb$ because   
 $\ds{ \sum_{\lambda_J < 0} } \lambda_J = -\tfrac{3}{4} < - \tfrac{1}{2} = - \tfrac{1}{d-1} $.

It is worth summarizing what is known about the EB subset of channels
constant on axes.
\begin{enumerate}
    \renewcommand{\labelenumi}{\theenumi}
    \renewcommand{\theenumi}{(\alph{enumi})}

\item  If $\Phi$ is EB, then  $\sum_J | \lambda_J | \leq 1$.

\item  If all $\lambda_J \geq 0$,  and $\sum_J | \lambda_J | \leq 1$, then $\Phi$ is EB.

\item   If all $\lambda_J \geq 0$ and some $\lambda_J = 0$, then $\Phi$ is EB.

\item  If $\Phi$ has one symmetry axix, then $\Phi$ is EB if and only if $\sum_J | \lambda_J | \leq 1$

\item  There are channels which satisfy $\sum_J | \lambda_J | \leq 1$, but are {\em not} EB.

\item  If all $\lambda_J \leq 0$, then $\Phi$ is EB.

\end{enumerate}     

We can also use the $\lambda_J$'s to state a necessary condition for an axis channel to be PPT:
\begin{thm} \label{thm:PPTd}
If a channel $\Phi$ constant on axes satisfies the PPT condition \linebreak
$({\cal T} \ot \Phi)(\proj{\beta}) \geq 0 $, then $\sum_J \lambda_J \le 1$.
\end{thm}
\prf   First observe that for any QC channel   
the antisymmetric subspace is in the kernel of
both  $(\id \ot \Phi^{\rm QC})(\proj{\beta})$ and
 $({ \cal T} \ot \Phi^{\rm QC})(\proj{\beta})$. 
To see this consider  $|v_{12} \ket  =  | f  \ot g  \ket - | g \ot f  \ket $ and
write $| f \ket = \sum_j x_j |j \ket$,  $| g \ket = \sum_j y_j |j \ket$ in the
basis corresponding to $\Phi^{\rm QC}$.   In this basis,
\bee
  (\id \ot \Phi^{\rm QC})(\proj{\beta}) =
  ({\cal T} \ot \Phi^{\rm QC})(\proj{\beta}) = \sum_{kk}  \proj{kk}
\eee 
and
$  | v_{12} \ket = \sum_{jk} ( x_j y_k - y_j x_k ) |jk\ket $
so that
 \bee
  (\id \ot \Phi^{\rm QC})(\proj{\beta})|v_{12} \ket  =
  ({ \cal T} \ot \Phi^{\rm QC})(\proj{\beta})| v_{12} \ket =  \td \sum_{kk}  | kk \ket (x_k y_k - y_k x_k) = 0
 \eee
 One similarly finds that the antisymmetric subspace is an eigenspace of $ ({ \cal T} \ot \id)(\proj{\beta})$
 with eigenvalue $-1$:
 \bee
 ({ \cal T} \ot \id)(\proj{\beta}) |v_{12} \ket &  = & 
    \td  \sum_{jk} |kj \kb jk |  \sum_{mn}( x_m y_n - y_n x_m ) |mn\ket \\
      & = &  \td \sum_{jk}     ( x_j y_k- y_k x_j )     |kj \ket  = ~ - |v_{12} \ket
 \eee
Thus,  if $\Phi$ has the form \eqref{axcdef}
 and satisfies the PPT condition, then
 choosing  $ |v_{12} \ket$ antisymmetric gives
 \be
     0 \leq    \bra v_{12}  |({ \cal T} \ot \Phi)(\proj{\beta})|v_{12} \ket  = - s
 \ee 
 which implies $s \leq 0$.   Then  \eqref{stcond} implies 
 $\sum_J \lambda_J = ds + 1 \leq  1$.  \qed

\subsection{EB and Bound Entanglement when $d = 3$}\label{Section:d3PPT}

We now consider some implications of the PPT and CCN conditions in more
detail when $d = 3$. Some of these results of this section
were obtained independently in \cite{BHN}.    However, they studied
the full set of states for ${\bf C}^3 \ot  {\bf C}^3$.   We consider only the state
representatives of channels constant on axes, which is a smaller set.
    In fact, the identity map $\id$ is the only
channel constant on axes whose CJ matrix is a pure state.   The other
extreme points  have CJ matrices with rank two.

For $d = 3$ it is shown in Appendix~\ref{sect:d3} shown that the maps
$\Psi_J^{\rm XEB}$ and $ \Psi_J^{\rm YEB}$ are extreme points of 
the convex subset of EB channels.    
These points lie in the ``base'' tetrahedron shown in Figure~\ref{fig:base},
which also shows the tetrahedron whose vertices $\Psi_J^{\rm XEB}$ are the four centers of the
faces.   Reflecting this small tetrahedron through its
center gives the convex hull of the four $ \Psi_J^{\rm YEB}$.   
The convex hull of the
eight points $\Psi_J^{\rm XEB}$ and $ \Psi_J^{\rm YEB}$ is a subset of the EB channels and is inscribed in 
the sphere  $\sum_J | \lambda_J | ^2 = \tfrac{1}{4}$,
which is precisely the set of PPT maps in the base tetrahedron. 
We conjecture that  all maps in this sphere are  EB; this is supported by 
numerical work of K. Audenaert  \cite{Aud}.

The observation about the sphere is a special case of the
following theorem which is proved in Appendix~\ref{sect:d3}:
\begin{thm}  \label{thm:PPT3}
When $d = 3$, a channel constant on axes is PPT if and only if it
satisfies  both $ \sum_J \lambda_J \leq 1$ and
\be   \label{ppt3}
   3  \sum_J \lambda_J^2 \leq  1 +  \sum_J \lambda_J +   \Big( \sum_J \lambda_J  \Big)^2  .
\ee
\end{thm}

We can use Theorem~\ref{thm:PPT3} to find examples of channels which
are PPT but not CNN.    Such channels are of some
interest because they correspond to bound entangled states.  
We first consider  $| \lambda_J| = x$ for all $J$.
The case all $\lambda_J =  x > 0$ is covered by Theorem~\ref{thm:EBT<1}
  and the case all $\lambda_J =  -x < 0$ has $x \leq \tfrac{1}{8}$ and is both
  PPT and  CCN.
Permutations of $[+x, -  x, - x, - x]$ have one symmetry
axis; it is shown in Section~\ref{sect:FHEB} that for these channels the
PPT and CCN regions always coincide.
 The only remaining possibility is permutations of
 $[+x,+x,-x,-x]$ for which $\sum_J   \lambda_J   = 0$
 and the CP condition \eqref{lamineq}    holds if and only if $x \leq \tfrac{1}{3}$.  
 In this case, \eqref{ppt3} becomes  $12 x^2 \leq 1$.    These we can conclude 
 that channels with multiplier $[+x,+x,-x,-x]$ are CP and bound entangled
 for  $\tfrac{1}{2 \sqrt{3}}  < x \leq \tfrac{1}{3}$.

\begin{figure}[here]  
 \begin{center}
 \includegraphics*[width=9cm,height=9cm,keepaspectratio=true]{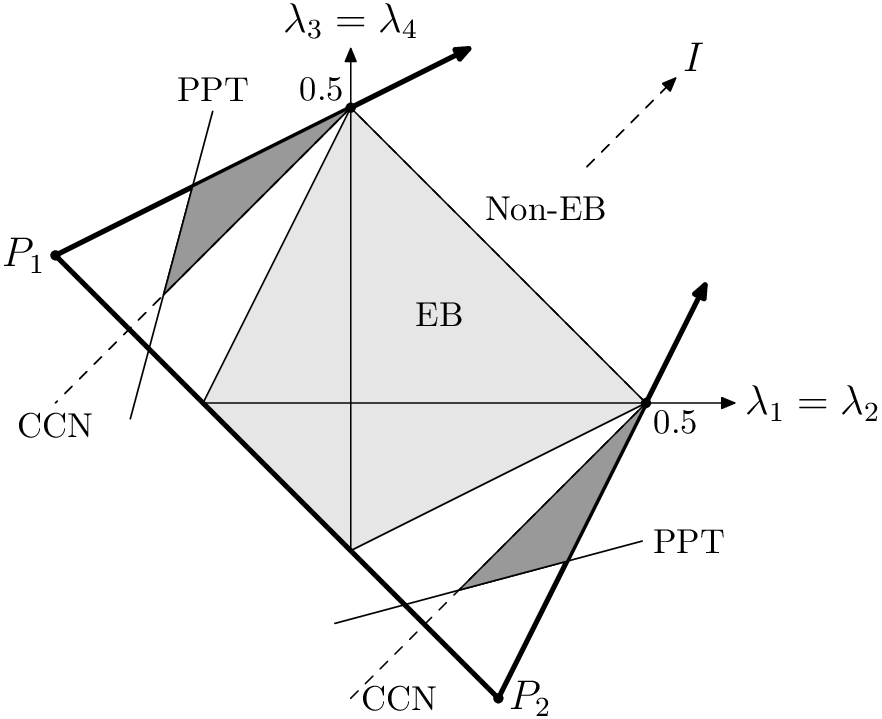}
\end{center}
  \caption{Qutrit channels with multiplier $[\lambda_1 ,\lambda_1, \lambda_3 , \lambda_3]$, which is the triangle $\id P_1  P_2$.  $P_1$ and $P_2$ correspond to the midpoints of two
 disjoint edges in the base tetrahedron of Figure~\ref{fig:base}.   
 Maps in the lightly shaded region are known to be EB; those in the
   dark regions correspond to bound entangled states.}  \label{fig:boundent}
\end{figure}

We now consider channels whose multipliers are permutations of
 $[x,x,-y,-y]$ with $x,y \geq 0$.      Let $S = \sum_j \lambda_J = 2(x-y)$ 
 and  $T = \sum_j | \lambda_J | = 2(x+y)$.   Then $x =\tfrac{1}{4}(T+S)$
 and $y =\tfrac{1}{4}(T-S)$.   The CP conditions \eqref{lamineq}
 become $S \geq - \half$
 and  $2x + y  = \tfrac{1}{4}(3T+S) \leq 1$.   The PPT condition \eqref{ppt3} 
 becomes
\bee 
    1 + S + S^2 \leq 6(x^2 + y^2) \leq  \tfrac{3}{4} (S^2 + T^2)
   \eee
or, equivalently, 
\be
      3 T^2 \leq (2 + S)^2
\ee
which is stronger than the CCN condition $T \leq 1$ when   $S+ 2 \leq \sqrt{3}$.  Thus
we can conclude that  channels of the form  $[x,x,-y,-y]$ give bound 
entangled states in the region
\be
    1 < T < \min \Big\{ \frac{2+S}{\sqrt{3}} , \frac{4 - S}{3} \Big\}
\ee
with   $\sqrt{3} - 2 < S < 1$.  In terms of $x,y$, this is the triangle bounded
by  the lines
\be
    2x + y = 1 , \quad    x+y = \half, \quad  \hbox{and}  
    \quad y = \tfrac{\sqrt{3} - 1}{2} + x(\sqrt{3}-2)
    \ee
    as shown in Figure~\ref{fig:boundent} with $\lambda_1=x, \lambda_3 = -y$.   If we drop the restriction that
    $\lambda_1 = \lambda_2$ and $\lambda_3 = \lambda_4$ one can find
    additional channels with bound entangled states for any value of
    $S \in \big(-\half, 1\big)$.
      
 \section{One symmetry axis}  \label{sect:FH}
 
 \subsection{General considerations}
 
 Channels of the form \eqref{gen} with exactly one $t_L$ non-zero have been
 considered by Fukuda and Holevo \cite{FH} who wrote them in  the form
 \be  \label{planar}
   \Phi(a,b)= b \id + a \Psi^\QC + (1-a-b) \cN.
 \ee
 When $d \pp 1$ MUB exist, assume they are labeled so that $\Psi^\QC$
 corresponds to  $J = 1$ so that  $\Phi$ has  multiplier $[a \pp b, b \ldots b]$.
 Even when a full set of MUB do not exist, \eqref{planar} is a well-defined channel
  with the QC map in the standard basis $|e_j\ket$.     Moreover,
  we can still associate $\Phi(a,b)$ with a multiplier
in the  generalized Pauli basis,  as described after \eqref{maprep}.  
Then $\phi_s = a + b$ when $V_s = Z^j$ for some $j$ and $\phi_s = b$
otherwise.
  These channels have exactly one symmetry axis, i.e., they satisfy
 the covariance condition
 $\Phi(U \rho U^\dag) = U \Phi(\rho) U^\dag$ when 
 $U = \sum_j  e^{i \xi_j} \proj{e_j} $.

  As observed in \cite{FH} these maps are CPT when
  $(a,b)$ is  in the convex hull of the points
 $(1,0),    (\tfrac{d}{d-1}, \tfrac{-1}{d-1}),  ( \tfrac{-1}{d-1} ,0) $
 as summarized in Table~1.      
The  CJ matrix for maps of the form \eqref{planar} can be written as
\be  \label{CJplan}
     \Gamma & = &   \td \sum_{jk} |e_j \kb e_k| \ot  \begin{cases}  \ds{ \sum_n }
          \big[ (a \pp b) \delta_{kn} + \td(1 \mm a \mm b) \big] \proj{e_n} & j = k \\
         \qquad    b  \, |e_j \kb e_k|  & j \neq k \end{cases} \\   \label{CJbeta}
       & = & \tfrac{1}{d^2} \Big[ (1-a-b) I + bd^2 \, \proj{\beta} + \sum_k ad \, \proj{e_k \ot e_k } \Big]  
\ee
with
 $| \beta \ket = \tfrac{1}{\sqrt{d}} \sum_k |e_k \ot e_k \ket $ maximally entangled.

\subsection{EB channels}  \label{sect:FHEB}

To find the subset of EB maps, observe that
the  PPT condition applied to \eqref{CJplan} is  
 $\pmx \td(1 \mm a \mm b) & b \\
   b &  \td(1 \mm a \mm b) \emx \geq 0$ or, equivalently, 
    $\td(1 \mm a \mm b)  \geq |b|$,
which can be written as
\bsq    \label{EBnec2} \be
      a + (d + 1)b &  \leq & 1    \qquad    b > 0  \\
        a - (d - 1)b & \leq & 1    \qquad   b < 0
\ee  \esq
When $a + b \geq 0$, \eqref{EBnec2} is equivalent to the CCN condition
$\sum_L |\lambda_L |\leq 1$.    This implies that for channels of the
form  $\eqref{CJplan}$ the PPT and CCN boundaries coincide.  

\begin{figure}[p]
 \begin{center}
  \vskip-1cm
 \hskip2cm  \includegraphics*[height=7.5cm,width=7.5cm,keepaspectratio=true]{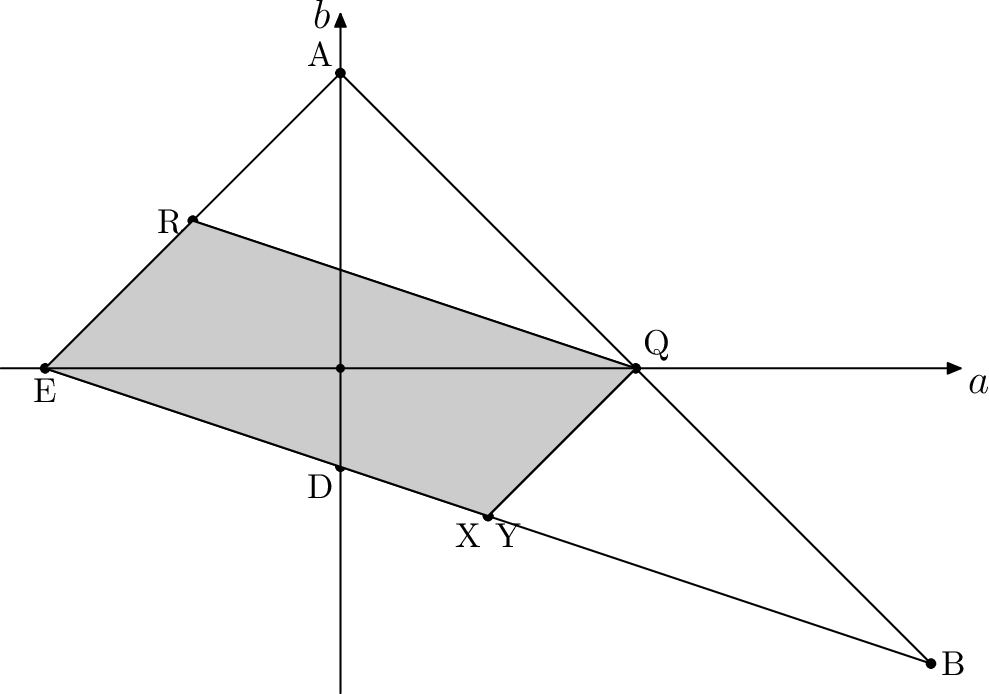}
  $\quad \qquad \begin{array}{c}  d = 2 \\  ~~  \\  ~~    \\  ~~  \\  ~~   ~~  \\  ~~  ~~  \\  ~~ 
    \\  ~~   ~~  \\  ~~  ~~  \\  ~~  \end{array}$ 
 \vskip-2cm
 \includegraphics*[height=12cm,width=12cm,keepaspectratio=true]{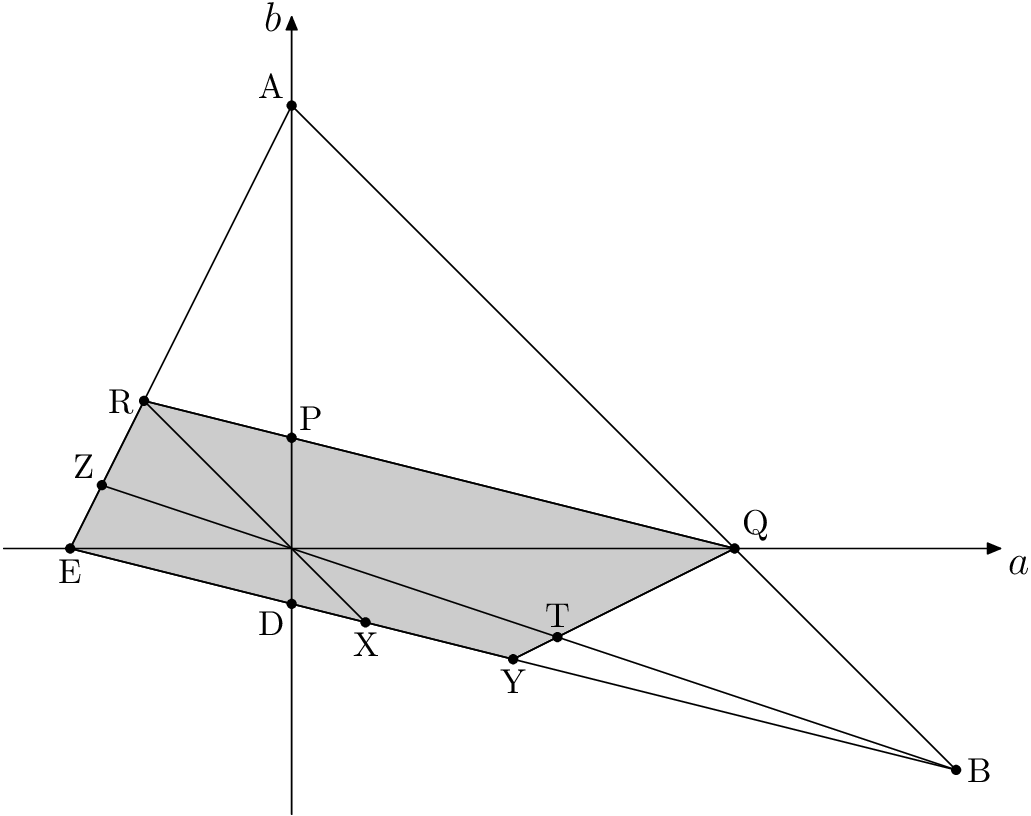}
  \vskip 0.4cm
 $\quad \qquad \begin{array}{c}  d > 2 \\  ~~  \\  ~~    \\  ~~  \\  ~~   ~~  \\  ~~  ~~  \\  ~~    \\  ~~  \\  ~~   ~~  \end{array}$ 
\vskip -2.5cm    See Tables~\ref{tab:FHCP},  \ref{tab:FHEB} and \ref{tab:FHother} for coordinates of marked points.
    
     \vskip0.4cm
  
   AB: $a + b = 1 \qquad $   AE: $a(d-1) - b = -1 \qquad $ BE: $ a + b(d+1) = - \frac{1}{d-1} \qquad$
  \vskip0.3cm
  
  RQ:  $a+b(d+1) = 1 \qquad$    YQ:  $a - (d-1) b = 1 $
  \vskip0.3cm
  
  RX:  $a+b = 0 \qquad$     ZB:  $a+db = 0 $
  
   \vskip0.4cm
  
  \caption{   Maps of the form  \eqref{planar} which are CPT are in the convex hull of ABE.  
Maps in the convex hull of ERQY are EB.  For $d=2$, the points $X$ and $Y$ coincide.} \label{fig:FH.EB}  
      \end{center}
  
 \end{figure}

When  $d  \pp 1$ MUB exist, one can  write the maps with $a = - b = - \td$ 
or $\tfrac{1}{d(d-1)}$  as 
$\ds{\sum_{J \neq L} \td \Psi_J^{\rm QC}}$   and  
$\ds{\sum_{J \neq L} \td \Psi_J^{\xeb}}$,  respectively, which implies that
they are EB.    Showing that they are EB for arbitrary $d$ is  
harder.\footnote{This problem was mentioned
at a talk in Torun in June, 2006.   Shortly after
this talk, J. Myrheim \cite{Myr} and P. Horodecki \cite{HorP2} independently gave explicit
constructions for separability of the CJ matrix for $b = \td$.   However, the
separability of the CJ matrix for $b = -\tfrac{1}{d(d-1)}$ was settled 
only by observing that X is on the line EY in Figure~\ref{fig:FH.EB} and Y is separable.}
 It is natural to conjecture that these
are also extreme points of the convex subset of EB maps of this type,
in which case the convex hull of ERQX in Figure~\ref{fig:FH.EB}
would give the EB channels. This is false, however.    The
next result says that all channels with one symmetry axis which satisfy the
PPT condition (or, equivalently, the CCN conditon) are  EB;
this corresponds to
the convex hull  of ERQY as shown in Figure~\ref{fig:FH.EB}.
\begin{thm}   \label{thm:EB=PPT}
A channel of the form \eqref{planar} is EB if and only if it
satisfies \eqref{EBnec2}.
\end{thm}
\prf  Since the set of EB channels is convex, it suffices to show that
each the channels corresponding to the points E, R, Q, Y in
Table~\ref{tab:FHEB} and Figure~\ref{fig:FH.EB} are EB.  
The points Q  corresponds to  $\Phi^{QC}$ which is EB and
E  has a separable CJ matrix  because $b = 0$.    Decompositions
showing that the CJ matrices for R and Y are separable are given
in Appendix~\ref{app:FHsep}.   \qed

    It is well known that the depolarizing channel, $\Psi^\dep_\lambda$,
     is EB for $\lambda \leq \tfrac{1}{d+1}$, which is consistent with
     Theorem~\ref{thm:EBT<1}.   If one ``depolarizes''  from  an
     extreme point  other than the identity, the resulting
channel    $\Phi = \lambda   \Psi_L^{\rmx} + (1 - \lambda) \cN $ has
one symmetry axis.   We can then use   Theorem~\ref{thm:EB=PPT} 
to  conclude that the channel is EB
when $\lambda \leq \tfrac{d-1}{2d-1}$, for which the limiting case
has multiplier $\big[ \tfrac{d-1}{2d-1} , \tfrac{-1}{2d-1}  , \ldots   \tfrac{-1}{2d-1}   \big]$.  
Note that the CP range is
$ \tfrac{ - d + 1}{d^2 -d +1} \leq  \lambda  \leq 1$ which has multiplier
$\big[    \tfrac{ - d + 1}{d^2 -d +1} , \tfrac{ 1}{d^2 -d +1} , \ldots, \tfrac{ 1}{d^2 -d +1}\big]$
at the boundary.   For $d = 3$ the EB portion of the line
segment $    \lambda   \Psi_1^{\rmx} + (1 - \lambda) \cN $ is bounded by
the channels with multipliers $  [\tfrac{2}{5}, \tfrac{-1}{5},  \tfrac{-1}{5} ,  \tfrac{-1}{5} ]  $
and $  [\tfrac{-2}{7}, \tfrac{1}{7},  \tfrac{1}{7} ,  \tfrac{1}{7}]    $.

\begin{table}[p]
$$  \begin{array}{|cclccc|}  \hline
& (a,b) & & & d = 2 & d = 3     \\[0.05cm]  \hline   
A&  (0,1) &  \id & [1,1, \ldots 1]  &[1,1,1] & [1,1,1,1]    \\[0.3cm]   
B &   (\tfrac{d}{d-1}, \tfrac{-1}{d-1})&  \Psi_{L}^{\rm X}  
  &[1, \tfrac{-1}{d-1} ,\ldots  \tfrac{-1}{d-1} ] & [1,-1,-1] &
  [1,  -\half ,   -\half  ,   -\half  ] \\[0.3cm]
E &   ( \tfrac{-1}{d-1} ,0) &  \Psi^\xeb_L  & [ \tfrac{-1}{d-1} ,0,\ldots 0]
    & [-1,0,0] & [  -\half , 0,0,0]  \\[0.3cm] \hline   
 \end{array}$$  \vskip-0.1cm
\caption{Extreme points of  CPT maps $b \id +a\Psi^\QC_1  +(1-a-b) \cN $.} \label{tab:FHCP}
\end{table}
 
\begin{table}[p]
$$  \begin{array}{|cccccc|}   \hline   
& (a,b) & & & d = 2 & d = 3   \\[0.05cm]     \hline  
Q &   (1,0) &  ~ \Psi_L^\QC & [1,0, \ldots 0]  & [1,0,0] & [1,0,0,0]  \\[0.35cm]
E &  ( \tfrac{-1}{d-1} ,0) &  \Psi^\xeb_L  & [ \tfrac{-1}{d-1} ,0,\ldots 0]  & [-1,0,0] & 
  [\frac{-1}{2}, 0,0,0]   \\[0.3cm]
R &  ( \tfrac{-1}{d} ,\td) & \ds{ \sum_{K \neq L} \td \Psi_L^\QC }
 & [ 0,\td, \ldots, \td] & [0, \half,\half] & [0,\tfrac{1}{3} , \tfrac{1}{3} , \tfrac{1}{3} ]  \\[0.35cm]
Y &  \left(\half,-\tfrac{1}{2(d-1)}\right)& \Psi^{\rm YEB}_L &
\big[\tfrac{d-2}{2(d-1)}, \tfrac{-1}{2(d-1)} ,\ldots  \tfrac{-1}{2(d-1)} \big]
     &[0, -\half , -\half] &  [ \tfrac{ 1}{4}, \tfrac{-1}{~4},  \tfrac{-1}{~4},\tfrac{-1}{~4}]   \\[0.35cm] \hline
 \end{array}$$ \vskip-0.2cm
\caption{Extreme points for  subset of  EB channels} \label{tab:FHEB}
\end{table}

\begin{table}[p]
$$  \begin{array}{|cccccc|}   \hline   
& (a,b) & & & d = 2 & d = 3   \\[0.05cm]     \hline  
X &  (\tfrac{1}{d(d-1)}, \tfrac{-1}{d(d-1)})& 
\ds{ \sum_{K \neq L} \td \Psi_L^\xeb }
 &[0, \tfrac{-1}{d(d-1)} ,\ldots  \tfrac{-1}{d(d-1)} ]
     &[0, -\half , -\half] &  [0, \tfrac{-1}{~6}, \tfrac{-1}{~6}, \tfrac{-1}{~6}]  \\[0.35cm]
     \cN& (0,0) &&  [ 0,0, \ldots 0] & [0,0,0] & [0,0,0,0]  \\[0.35cm]
         P & (0,  \tfrac{1}{d+1}) & \Psi^\dep_{\frac{1}{d+1}}& &[\tfrac{1}{3} , \tfrac{1}{3} , \tfrac{1}{3} ] & [\tfrac{1}{4} ,\tfrac{1}{4} ,\tfrac{1}{4} ,\tfrac{1}{4}  ]
  \\[0.35cm]
 D & (0,  \tfrac{-1}{d^2-1}) & \Psi^\dep_{\frac{1}{d^2-1}}& &[-\tfrac{1}{3} ,- \tfrac{1}{3} , -\tfrac{1}{3} ] & [-\tfrac{1}{8} ,-\tfrac{1}{8} ,-\tfrac{1}{8} ,-\tfrac{1}{8} ]
     \\[0.35cm] 
      T & (\tfrac{d}{2d-1},\tfrac{-1}{2d-1}) & & &[\tfrac{1}{3} ,  \tfrac{-1}{3} , \tfrac{-1}{3} ] & 
    [\tfrac{2}{5}, \tfrac{-1}{5},  \tfrac{-1}{5} ,  \tfrac{-1}{5} ]    \\[0.35cm] 
      Z & (\tfrac{-d}{d^2-d+1} ,  \tfrac{1}{d^2-d+1} ) &  & &[\tfrac{-1}{3} ,  \tfrac{1}{3} , \tfrac{1}{3} ] & 
     [\tfrac{-2}{7}, \tfrac{1}{7},  \tfrac{1}{7} ,  \tfrac{1}{7}]   \\[0.35cm]    
 \hline
 \end{array}$$ \vskip-0.2cm
\caption{Other interesting points} \label{tab:FHother}
\end{table}

 \subsection{Multiplicativity}\label{sect:FHmult}
 
 Although this topic is more fully studied in the next section,
 where complete definitions are given, it is worth making some
 observations here.    We use ``multiplicative'' to mean that
 \eqref{mult} holds with $\Omega$ arbitrary.
One can apply Fukuda's  lemma \cite{F1} to show that 
\be
        \Phi = \Psi^{\rm dep}_y \circ \Psi^\PD_{L,x} =  xy \id + (1-x)y \Psi^\QC_L + (1-y) \cN
\ee
is multiplicative for $\tfrac{-1}{d-1} \leq x \leq 1$ and
$\tfrac{-1}{d^2-1} \leq y \leq 1$.    The relations $b = xy$ and $a = (1-x)y$,
 imply that
\be
     \tfrac{-1}{d^2-1} \leq a + b  \leq 1  \quad \text{and}  \quad   \tfrac{-1}{d-1} \leq  \frac{b}{a+b} \leq 1.
\ee

This gives the following result.
\begin{thm}  \label{thm:fuk}
A map of the form \eqref{planar} is multiplicative if either of the following sets of conditions hold

~i)  $a  > 0$ and $a+bd \geq  0$, or

ii)   $a < 0$ and $-b -  \tfrac{1}{d^2-1} \leq a \leq - bd$.
\end{thm}
The second set of conditions (ii) corresponds to a very small region
entirely contained within the set of EB channels.
\begin{figure} 
\begin{center}
\vskip2cm \includegraphics*[height=6 cm,width=7.5cm,keepaspectratio=true]{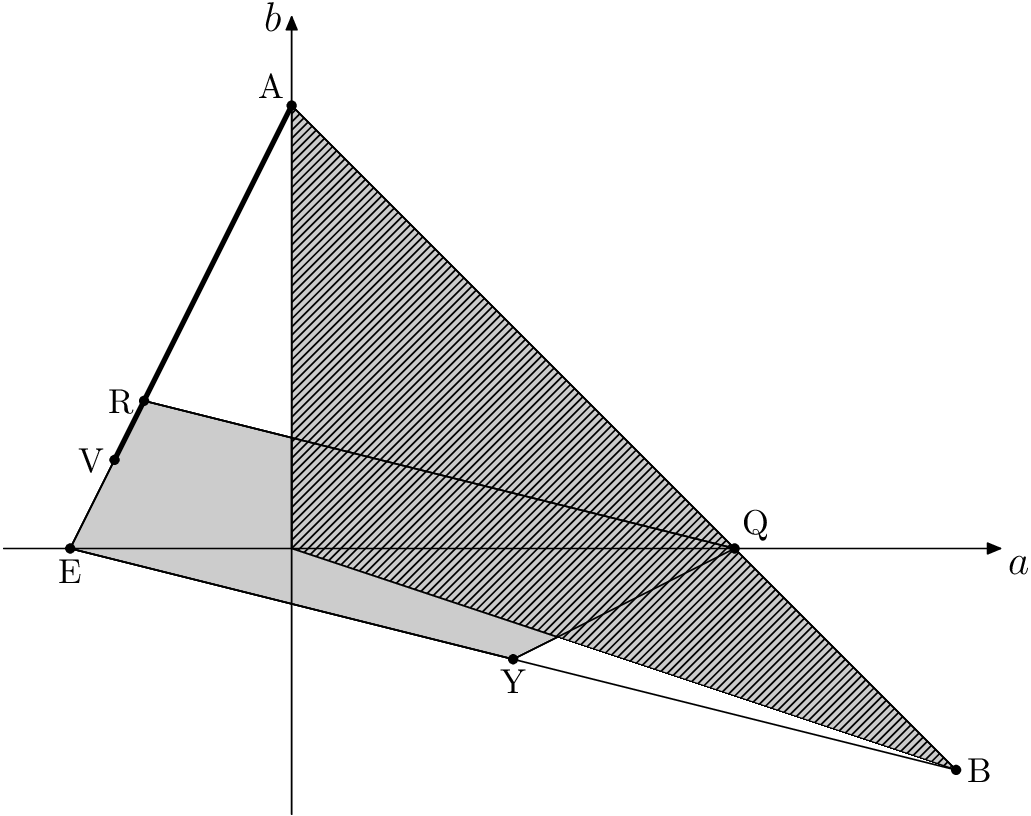}
\label{fig:FHadditive}
\caption{Channels with one symmetry axis for which multiplicativity
holds for tensor products with arbitrary channels. The shaded triangle $AB{\cal N}$ shows channels implied by Theorem \ref{thm:fuk}.} 
\end{center}
\end{figure}

The points on the line segment   AV in Figure 5  
correspond to
maximally squashed channels. For any fixed $(a_*, b_*)$ on the segment AV, 
if one can show that both \eqref{mult} holds and 
$\nu_p\big(\Phi(a_*,b_*)\big) = \nu_p\big(\Phi(0,b_*)\big)$, then it follows from 
Theorem~\ref{thm:star}
that multiplicativity holds for all $\Phi(a,b_*)$ with $a_* \le a \le 
0$.
Thus, the multiplicativity  problem
for the triangle  $R, A, (0,\td) $ is reduced to the line AR.

Proving
multiplicativity for the triangle  YTB presents a different challenge.

\section{Optimal output purity}  \label{sect:mult}

\subsection{General considerations}  \label{sect:pmult}

One measure of optimal output purity is the minimal output entropy, 
defined as
$S_{\min}(\Phi) = \inf_\gamma  S[\Phi(\gamma)]$ where $S(\rho) = - \tr 
\rho \log \rho$
denotes the entropy of a quantum state.   The additivity conjecture is 
\be   \label{add}
   S_{\min}(\Phi \ot \Omega) = S_{\min}(\Phi) + S_{\min}(\Omega).
\ee
This conjecture is particularly important because Shor \cite{Shor2} has 
shown that
it is globally equivalent to several other important conjectures: 
additivity of
the Holevo  capacity, additivity of the entanglement of formation under 
tensor
products, and superadditivity of entanglement of formation.   Recently,
Shirokov \cite{Shir} showed that if \eqref{add} holds for all channels 
$\Phi:M_d \mapsto M_d$,
then this collection of additivity conjectures also holds in infinite 
dimensions.

Another measure of the optimal output purity  of a channel is the
maximal output $p$-norm defined as
\be
\nu_p(\Phi) = \sup_{\gamma} \norm{\Phi(\gamma)}_p =  \sup_{\gamma}  
\big( \tr [\Phi(\gamma)]^p \big)^{1/p}
\ee
It has been conjectured \cite{AHW,KR3} that $\nu_p(\Phi)$ is {\em 
multiplicative} in the
sense
\be   \label{mult}
\nu_p(\Phi \ot \Omega) =  \nu_p(\Phi)  \, \nu_p(\Omega)
\ee at least for $1 \leq p \leq 2$.    Moreover,
it was shown in \cite{AHW} that if
\eqref{mult} holds for $p \in (1,1+\epsilon)$ for some $\epsilon > 0$, 
then
\eqref{add} holds.

Recently, Fukuda \cite{F2} showed that if
\eqref{add} or \eqref{mult} holds for all channels
$\Phi: M_{d_1} \mapsto M_{d_2} $ which map
$\tfrac{1}{d_1} I_{d_1}$ to $\tfrac{1}{d_2} I_{d_2} $
then it holds for arbitrary $\Phi$.   Thus the general
case is reduced to a kind of extended unital channel.
(When $d_1 \neq d_2$ a CP map cannot be both TP and unital.)
Our original motivation for studying  channels constant on axes was to 
find a class to which
one could extend King's proof \cite{King2} of this conjecture in the 
case of unital qubit channels.
Instead, we have merely gained additional insight into the  reasons his
argument does not work when $d > 2$.   We can however prove 
multiplicativity
for channels constant on axes in the important case $p = 2$, as is 
shown
Section~\ref{sect:p2mult}.

We conjecture that for channels constant on axes, the maximal output
$p$-norm and minimal output entropy are both achieved with an axis
state.
\begin{conj} \label{conj:axisachieve}
Let $\Phi$ be a channel  of the form \eqref{axcdef}.
Then the maximal output $p$-norm and minimal output entropy can
be achieved with an axis state, i.e., for each $p$ one can find $L$ 
(which may depend upon $p$)   such that
  $\nu_p(\Phi) =  \norm{ \Phi(\proj{ \psi_n^L}) }_p = 
\norm{\Psi^{\dep}_{\lambda_L}}_p$,
and $S_{\min}(\Phi) = S\big( \proj{ \psi_n^L} \big) = 
S_{\min}(\Psi^{\dep}_{\lambda_L})$
for some $L$.
\end{conj}
When some of the $\lambda_J < 0$, the axis $L$ can depend upon $p$ and
our evidence for this conjecture is only numerical, as described in the 
next
section.    When all $\lambda_J \geq 0$, the analogy with unital qubit 
channels
suggests that $L$ satisfies $\sup_J \lambda_J = \lambda_L$ and it is 
easy to
see that if $\lambda_J < \lambda_L$ then $ \norm{ \Phi(\proj{ 
\psi_J^n}) }_p \leq  \norm{ \Phi(\proj{ \psi_L^n}) }_p $ for all
$p$.   However, we have not been able to exclude the possibility that 
$\nu_p(\Phi)$ is attained on a superposition of axis states.  

Conjecture~\ref{conj:axisachieve} is known for certain classes of axis channels, such as QC channels and depolarizing channels, and it is shown to hold for all axis channels in the special cases $p=2$ and $p=\infty$, as discussed in Appendix \ref{app:mult}.

The following result, which is
a special case of Lemma~\ref{lemma:crit} proved in 
Appendix~\ref{app:mult},
is consistent with this conjecture.   \begin{thm}   \label{thm:crit}
   Let $\Phi$ be a channel constant on axes and $\gamma(t)$ a 
differentiable
   one-parameter family of pure states with $\gamma(0) =  
\proj{\psi_n^J}$
   an axis state.  Then for all $p$, $ \nrm\Phi[\gamma(t)]\nrm_p$ has a 
critical point at $t = 0$.
   Moreover, $S_{\min}\big(\Phi[\gamma(t)]\big)$ also has a critical 
point at $t = 0$.
   If $\Phi_1$  and $\Phi_2$ are both constant on axes, this extends to 
inputs
    $\gamma_{12}(t)$ with $\gamma_{12}(0) =  \proj{ \psi_{n}^{J} \ot 
\psi_{m}^{K} }  $.
\end{thm}

Whenever  $\nu_p(\Phi)$ is achieved with an axis state, one can use
Fukuda's lemma \cite{F1} to show that if \eqref{mult} holds for $\Phi$, 
it can be
extended to $\Phi \circ M_J$, where $M_J$ is a convex combination of
conjugations on a single axis as in \eqref{Mdef}.     The key point is 
that
$M_J(\proj{\psi_n^J}) = \proj{\psi_n^J}$.
\begin{thm}
Let $\Phi$ be a channel constant on axes and let $M_J$ be given by 
\eqref{Mdef}.
If $\nu_p(\Phi) = \norm{ \Phi(\proj{\psi_n^J})}_p$ and \eqref{mult} 
holds for
$\nu_p (\Phi \ot \Omega)$, then it also holds for $\Phi \circ M_J$.
\end{thm}

\subsection{Numerical study of new behavior}
  
  For a channel constant on axes, the output of any axis state is
\be   \label{nubd}
   \Phi(\proj{\psi^{L}_n}) =  (1 - \lambda_L) \td \rmi + \lambda_L    \proj{\psi^{L}_n} 
\ee
which has eigenvalues $\td[1+(d \mm 1) \lambda_L ] $ and $\td (1 - \lambda_L)$ with
degeneracy $d \mm 1$.      This implies  $\nu_p(\Phi) \geq \nu_p(\Psi^{\dep}_{\lambda_L})$
when $\lambda_L \geq - \tfrac{1}{d^2 -1}$.
When all $\lambda_K \geq 0$, we conjecture that this is optimal, i.e.,
$\nu_p$ is achieved with an axis state corresponding to the largest $\lambda_K$
and provide some evidence in this direction.     We also show that channels
with negative mutipliers can have fundamentally different behavior.

In particular,  it can happen that
$0 < - \lambda_{L_1} = |\lambda_{L_1} | < \lambda_{L_2} $ but
$\norm{  \Phi(\proj{\psi^{L_1}_n}) }_p >  \norm{  \Phi(\proj{\psi^{L_2}_n}) }_p$.
For example, consider the channel 
\be    \label{posneg}
   \Phi = a \Psi_1^{\rm X} +  (1-a)\Psi_2^{\rm X}  ~~ \text{with multiplier} ~~
       [\tfrac{3a-1}{2},  \tfrac{2-3a}{2} ,-\half,-\half] 
\ee
For $a = \tfrac{2}{3}$ this becomes $[+0.5,  0 ,-0.5, -0.5]$  which has larger
output $p$-norms when $1 < p < 2$ for inputs along the $-0.5$ axes than for 
those along the $+0.5$ axis.   This behavior persists for  
  $[0.6, -0.1, -0.5, -0.5]$ when $1 < p < 1.2$.    Moreover,
for $0 < \lambda_1 < 0.65$,
the minimal output entropy of the channel with multiplier  \linebreak
$[\lambda_1, 0.5 - \lambda_1, -0.5, -0.5]$
is $S_{\min}(\Phi) =  S\big[ \Phi(\proj{\psi^3_n})\big] = 1$, but
 inputs on the ``long''  axis have $S\big[ \Phi(\proj{\psi^1_n})\big] > 1.$

Numerical studies of the minimal output entropy of channels with
multiplier $[\lambda_1, 0.5 - \lambda_1, -0.5, -0.5]$ have been carried out for
$\lambda_1$ near the crossing point $\lambda_1 = 0.659$.  For a single
use of the channel,  $S_{\min}(\Phi)$ is always achieved with an axis state
and satisfies $S_{\min}(\Phi) = 1$ for $\lambda_1 < 0.659$.   For the
product, one finds  $S_{\min}(\Phi \ot \Phi) = 2$ is always achieved with a
product of axis states.   Moreover, near the crossing the maximally 
entangled state  $| \beta \ket = \tfrac{1}{\sqrt{d}} \sum_n | \psi_n^1 \ot \psi_n^3 \ket$
has entropy  $S[ (\Phi \ot \Phi)(\proj{\beta})] = 2.74041$.

\subsection{Non-negative multipliers}  \label{sect:posmult}

King's approach to the unital qubit channels is to reduce the problem 
to multiplicativity
of ``two-Pauli'' channels''   $\Psi_{L,x}^{\rm MxSq}$ by considering 
channels of the form
$[\lambda_1, \lambda_2, x]$ with $| \lambda_j| \leq x$.   This subclass 
of channels
has extreme points  with multipliers
\be  \label{king}
[x,x,x], [-x,-x,x], [2x \mm 1,x,x], [x,2x \mm 1,x], [1 \mm 2x,-x,x], 
[-x,1 \mm 2x,x]
\ee
for $x > \tfrac{1}{3}$.    Here, $\Psi_{J,x}^{\rm MxSq}  $ has 
multiplier $ [2x-1,x,x]$.
If  one can show that  $\nu_p(\Psi_{J,x}^{\rm MxSq} ) = 
\nu_p(\Psi_x^{\dep})$, and
that  $\nu_p(\Psi_{J,x}^{\rm MxSq}  \ot \Omega) $ satisfies 
\eqref{mult}
then  multiplicativity follows from Lemma~\ref{lemma:elem}, first using
$B = \nu_p(\Psi_x^{\dep})$ and then using $B = \nu_p(\Psi_x^{\dep}) \,  
\nu_p(\Omega)$.
King's argument exploits the fact that changing $\lambda_j \raw - 
\lambda_j$ for
$j = 1,2$ is equivalent to a unitary conjugation with $\sigma_z$.   
This property does not extend to channels constant on axes.  However, 
we can
make an analogous reduction on the subset of channels with non-negative 
multipliers
under the assumption that Conjecture~\ref{conj:axisachieve} holds for 
these channels. 

\begin{figure}[p]
    \vskip-2cm    \begin{center}
       
\includegraphics*[height=5cm,width=5cm,keepaspectratio=true]{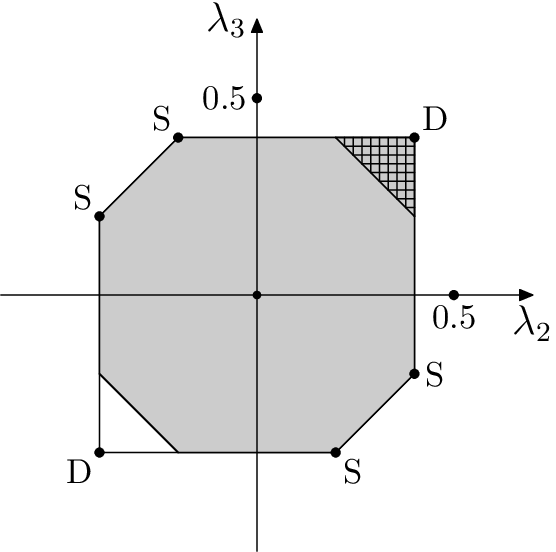}
\hskip 2 cm        
\includegraphics*[height=5cm,width=5cm,keepaspectratio=true]{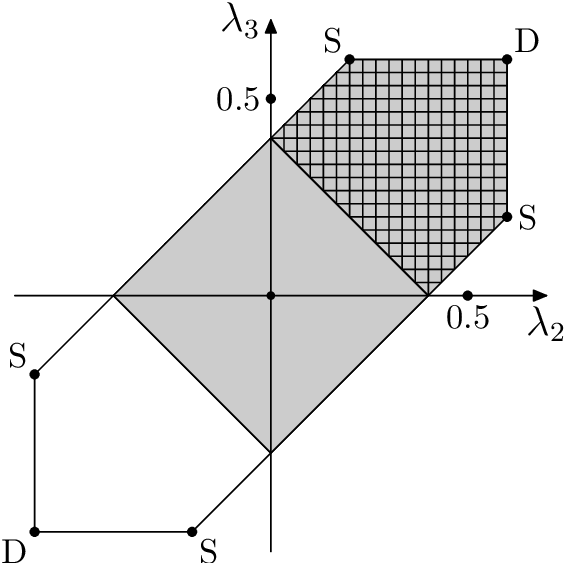}

\hskip-1.2cm $ \tfrac{1}{3}  \le x \le \tfrac{1}{2}   $  \hskip5.6cm $  
  x  >  \tfrac{1}{2} $
\end{center}
   \caption{Qubit channels with  $ |\lambda_j| \leq  x = \lambda_1$, 
with grey
    shading  indicating the EB region and  cross-hatch
    the subset of the non-EB channels with all $\lambda_j \geq 0$.  The
    evident symmetry for $\lambda_j \mapsto - \lambda_j$ is lost for $d 
 > 2$ but the
    picture for $\lambda_j \geq 0$ is similar. }
    \label{fig:slice}
    \end{figure}

  For qubits, the subset of channels with multiplier
$[\lambda_1, \lambda_2, x]$ with $0 \leq  \lambda_j \leq  x$  has
extreme points
\bee
[0,0,x],  \quad   [0,x,x],  \quad  [x,0,x], \quad  [x,x,x]  \qquad  &  
&   ~ x \leq  \half  \\       ~  [0,0,x],  [0,1 \mm  x, x], ~   [1 \mm  
x, 0,x], ~ [2x \mm  1,x,x], ~  [x,2x \mm  1,x],  [x,x,x]         &  &  
~ x  > \half
  \eee
  as shown in Figure~\ref{fig:slice}.
In both cases, the first 3 channels are EB and the last
the depolarizing channel.   The difference between the two situations 
is
that the latter includes channels of the form $\Psi_{J,x}^{\rm MxSq}  $ 
but the
former does not.   For $d > 2$, the convex set of channels which are 
not EB and satisfy
  $0 \leq \lambda_J \leq \lambda_{d+1} = x$ has analogous extreme 
points.
  We observe here only that any channel   $\Phi_x$ in this set is a 
convex
  combination of
\begin{itemize}

\item[a)]      $ \Psi^{\rm EB}_x$ and $  \Psi_x^{\dep}$ ~~ when  $0 <  
x \leq \td $, and
     \item[b)]          $ \Psi^{\rm EB}_x$,  $  \Psi_x^{\dep}$ and 
$\{\Psi_{J,x}^{\rm MxSq}:    J =1,2 \ldots d \} $  ~~ when $  \td < x 
\leq  1$
\end{itemize}
  where $ \Psi^{\rm EB}_x$ denotes  some EB channel whose multiplier
  satisfies $0 \leq \lambda_J \leq x$.
  The extreme EB channels of this type will  be permutations of
     $[0, \ldots, 0, 1 \mm  \kappa x, x, \ldots x]$ with $\kappa$ 
chosen
     so that $0 < 1 -  \kappa x \le x$.   
     
    \begin{figure}[p]       \vskip-1cm
    \begin{center}
    \vskip2cm
   
\includegraphics*[height=5cm,width=5cm,keepaspectratio=true]{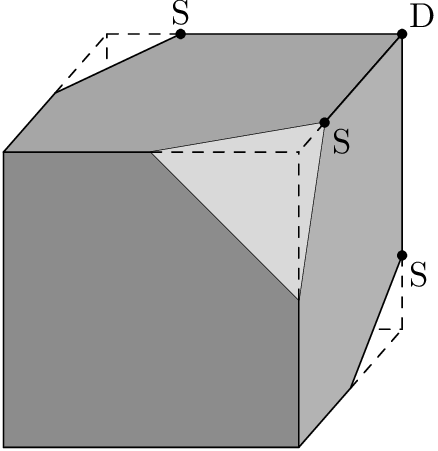}
       \hskip2cm     
\includegraphics*[height=4cm,width=4cm,keepaspectratio=true]{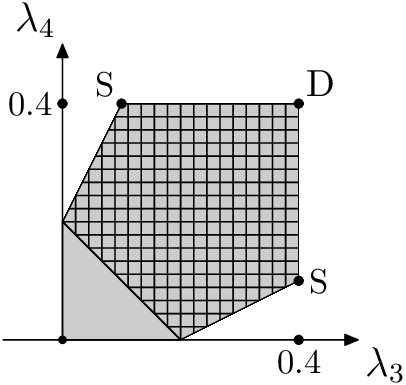}

     \end{center}

        \caption{Qutrit channels with $0 \leq \lambda_j \leq \lambda_1 
= 0.4$. The 3-dimensional
        view is a cube with three corners removed. The verticies of the 
front corner
        are the squashed channel
         $S = [0.4,0.4,0.4,.1]$ along with $[0.4, 0.4, 0.2, 0]$ and 
$[0.4, 0.2, 0.4, 0]$.          The two-dimensional view shows the face 
of the cube with $\lambda_2 = 0.4$.}
    \label{fig:slice3d}
    \end{figure}
            
            The following consequence of Lemma~\ref{lemma:elem} shows that if 
Conjecture~\ref{conj:axisachieve} holds for EB channels with
non-negative multipliers,
then we can reduce the general situation to the maximally squashed 
channels.
\begin{thm}\label{thm: multconj}
Let $\Phi$ be a channel of the form \eqref{multform}  with  $\lambda_J 
\geq 0$ for all $J$.
Choose $L^*$ so that $\sup_J \lambda_J = \lambda_L^*$.Then

a)  If   Conjecture~\ref{conj:axisachieve} holds  for $\Phi^{\rm 
MxSq}_{J,\lambda_L*}$ and any EB channel with  $0 \leq \lambda_J \leq 
\lambda_L^*$, then
\be  \nu_p(\Phi) =  \norm{  \Phi(\proj{\psi^{L^*}_n}) }_p = 
\nu_p(\Psi^{\dep}_{\lambda_{L^*}}) \ee

  b) If  $p \geq 1$, the hypotheses in (a)  above hold, and
\eqref{mult} holds for $\nu_p(\Phi^{\rm MxSq}_{J,\lambda_L*} \ot 
\Omega)$,
then  \eqref{mult}  holds for $\nu_p(\Phi \ot \Omega)$.
           \end{thm}

\prf  A channel satisfying the hypothesis above is a convex
combination of $\Psi^{\dep}_x$ and   EB channels  with
multipliers that are permutations of  $[0, \ldots, 0, 1 \mm  \kappa x, x, \ldots x]$.  By the 
remark in Section~\ref{sect:FHmult} such
EB channels have the same maximal output p-norm
as  $\nu_p(\Psi^{\dep}_x)$.  Then part (a) follows   from 
Lemma~\ref{lemma:elem} with
$B = \nu_p(\Psi^{\dep}_x)$ and the fact that the bound can be
attained with an axis state.     Both depolarizing channels 
\cite{King3} and
EB  channels  \cite{King4} are multiplicative for all $p$.   Therefore, 
part (b) also  follows from the Lemma~\ref{lemma:elem} with
$B =  \nu_p(\Psi^{\dep}_x) \, \nu_p(\Omega)$.  \qed

In the case of qubits, Conjecture~\ref{conj:axisachieve} is known to hold, so Theorem \ref{thm: multconj} gives an new proof of multiplicativity
for channels whose multiplier $[\lambda_1, \lambda_2, \lambda_3]$
satisfies  $0 \leq \lambda_j \leq \half$ and $\sum_j \lambda_j > 1$.
One can then conjugate with $\sigma_j$ and combine with known
results about EB channels to prove that \eqref{mult} holds
for any unital qubit channel  with all $| \lambda_j| \leq \half$.
This last step does not extend to $d > 2$. If the maximal $\lambda_j$ is greater than $\half$ for a qubit channel, this gives a new reduction of multiplicativity to maximally squashed channels. Both cases for qubits can be seen in Figure \ref{fig:slice}.

     Recall  $\Phi^{\rm MxSq}_{L,x} = \sum_{K \neq L}  
\Psi_{L,\zeta}^{\PD} $ with
$\zeta$   as in \eqref{AD}   and that  intuitively
one would expect that  $\nu_p( \Phi^{\rm MxSq}_{L,x})$ is achieved with
a state which is $0$ on the ``short'' axis $L$, i.e., for which $u_{Lj} 
= 0$ in
\eqref{dbloch2}.   For such states, King's proof  \cite{King3} of
the multiplicativity of the depolarizing channel carries over.  
However, the
channel $\Phi^{\rm MxSq}_{L,x} $ has symmetry {\em around} the axis L.   
This
does not allows  one to assume  that $w_{Lj} = 0$.
To overcome this problem, King \cite{King2} rewrites the anti-damping 
channel  in
the convex form
$\Phi^{\rm MxSq}_{3,x}  = \frac{3x-1}{2x} \Psi_{2.x}^{\PD} + 
\frac{1-x}{2x} \sigma_1 \Psi_{2.x}^{\rm MxSq} \sigma_1$
or, equivalently,
\be   \label{kingq}
    [x,x,2x-1] = \tfrac{3x-1}{2x} [x,1,x] + \tfrac{1-x}{2x} [x,1-2x,-x]
\ee
after making a rotation so that  $\rho = \half( I + w_1 \sigma  + w_3 
\sigma_3)$,
i.e., $w_2 = 0$.   Then both channels on the right have $\nu_p(\Psi) = 
\nu_p(\Phi_x^{\dep})$.   However, the channel
$[x,1-2x,-x]$ with negative multipliers has no direct analogue in the 
$d > 2$ case.
Therefore, King's argument does not generalize. Amosov \cite{Am} has 
given a new proof of additivity \eqref{add} for
unital qubit channels.   Because his argument is based on King's
decomposition \eqref{kingq}, it does not readily generalize to $d > 2$.

Proving $\eqref{mult}$ for channels  with positive
multipliers seems to require a new approach to the multiplicativity for
channels $\Phi^{\rm MxSq}_{L,x} $.    However, we have reduced the
problem to this case.  Channels with some
negative multipliers present a different challenge.

\subsection{Results for $p = 2$}  \label{sect:p2mult}

In the case $p = 2$, we can prove more, including multiplicativity
for all channels constant on axes.  
The results of this section are based on the following theorem
which was proved by Fukuda and Holevo in \cite{FH} in the case
of generalized Pauli matrices; inequality \eqref{FKNup}  was obtained 
independently by  Nathanson in \cite{KMNR}.   A proof, which
 is essentially identical to that in \cite{FH}, is presented in 
 Appendix~\ref{app:FHN} for completeness.

 \begin{thm}   \label{thm:FHN} {\em (Fukuda-Holevo-Nathanson)}
Let $\Phi$ be a CPT map which is diagonal when represented in
an OBU, and let  $\phi_s$ denote its diagonal
elements.  Then
\be   \label{FKNup}
    \nu_2(\Phi) \leq \Big( \td \big[ 1 + (d-1) \sup_s | \phi_s|^2 \big] \Big)^{1/2}.
\ee 
Moreover, if the bound \eqref{FKNup} is attained, then
\be
 \nu_2(\Phi \ot \Omega) =  \nu_2(\Phi)  \, \nu_2(\Omega)
\ee
for any CPT map $\Omega$.
\end{thm}

This result implies that  all  channels  constant on axes
 satisfy the multiplicativity conjecture for $p = 2$.
\begin{thm}
Let  $d$ be a prime power and $\Phi: M_d \mapsto M_d$ be  a CPT 
map constant on axes.    Then  \eqref{mult} holds for $p = 2$, i.e., 
$    \nu_2(\Phi \ot \Omega) =  \nu_2(\Phi)  \, \nu_2(\Omega) $
with $\Omega$  an arbitrary CPT map.
\end{thm}
\prf  By Theorem~\ref{thm:FHN}, it suffices to show that
\be   \label{p2hold}
    [\nu_2(\Phi)]^2 =  \td \big[1 + (d-1) \lambda^2 \big]
\ee
where $\lambda = \sup_L |\lambda_L | = \sup_L |s + t_L | $.   
For channels constant on axes, it is straightforward to verify that
\eqref{p2hold} is attained  with any axis state $\proj{\psi^{L^*}_n}$ for which
$|\lambda_{L^*}| = \lambda$.   \qed

One can extend this slightly to cover channels of the form \eqref{gen},
which includes Example 4 in \cite{FH}.
\begin{thm}
Let  $\Phi$ be  a channel on $M_d$ of the form \eqref{gen} with
 $\kappa$ MUB.
Then \eqref{mult} holds for $p = 2$ with $\Omega$  an arbitrary CPT map.
\end{thm}

For channels in the OBU $\{ U_{Lj} \}$,
 one can relax the requirement
that the channel is ``constant'' on the longest axis.
\begin{thm}
Let  $\Psi$ be a channel constant on axes,
  $ \lambda_{L_0}  \geq |\lambda_L |  ~ \forall ~ L$,
and $M_J$  a channel of the form \eqref{Mdef} with $J = L_0$.
If either
$    \Phi = x M_J + (1-x) \Psi $ , with  $0 \leq x \leq 1$, or
$     \Phi =  \Psi \circ M_J $,
then \eqref{mult} holds for $p = 2$ with $\Omega$  an arbitrary CPT map.
\end{thm}
 \prf  Both $\Psi$  and  $M_J$   are diagonal in the orthogonal unitary basis
  $\{ U_{Lj} \}$ with multipliers $\lambda_L$ and  $ \mu_{Lj}  $ respectively.
In both cases, one can verify  that $\Phi$ is also diagonal
with $\phi_{L_0 j} = \lambda^*$ independent of $j$ and 
$\phi_{L_0 j}  \geq |\phi_{Lj}|$.   (In the first case,
$\lambda^* =  x + (1-x)  \lambda_{L_0} $; in the second
$\lambda^* =  \lambda_{L_0} $.)
Therefore, $\sup_s |\phi_s| =  \lambda^*$  and any axis state 
 $\proj{\psi^{L_0}_n}$ saturates the bound \eqref{FKNup}.  The
 result then follows from Theorem~\ref{thm:FHN}.    \qed

The  channel $\Phi$ is constant on the ``longest'' axis in the sense 
that the multiplier $\phi_{L_0j}$ is independent of $j$ on this axis.   But it is
the constraint $a_{L_0 j}$ independent of $j$ that has been relaxed.     
 Maps of the form above with $\Psi$  a depolarizing channel were studied in
\cite{DR} and shown to satisfy \eqref{mult} for all $p$.

\bigskip


\section{Bloch sphere picture}

The Bloch sphere picture has proved so useful for $d = 2$ that there have
been numerous attempts to extend it to higher dimensions, and 
\eqref{rhorep} can be regarded as such an extension.   Moreover, the
conditions \eqref{rho.els} extend the standard criterion on the components
of the vector representing a density matrix.   The fact that the vector in \eqref{rhorep}
are complex rather than real is an inessential consequence of our decision 
to focus on OBU rather than bases with Hermitian elements.
(Replacing $W_J^j$ and  $W_J^{-j}$ by  $\half(W_J^j + W_J^{-j})$
and $\tfrac{i}{2}(W_J^j - W_J^{-j})$, replaces $v_{Jj}$ and $v_{J,d-j} $ by 
${\rm  Re}  \, v_{Jj} $ and ${\rm Im}  \, v_{Jj} $ respectively.)
The essential problem is that \eqref{rho.els}
is a necessary, but not sufficient condition for a matrix of the form \eqref{rhorep} 
to yield a positive semi-definte matrix.   Finding simple sufficient conditions for
positivity, or even purity,  is the real roadblock.

For qubits, all vectors on the surface of the unit ball correspond to pure states
and its image under a CP is an ellipsoid contained in this ball.   As shown 
in \cite{FA,KR1,RSW} not every ellipsoid corresponds to a CP map, but those
that do define a unique CP map with positive multipliers.    However, the role
of negative multipliers is lost completely.     The Bloch sphere picture does not
show rotations (unless composed with another map) and does not show the
effect of, e.g.,  a bit flip even when composed with another map.

The channels presented here do allow a partial generalization of the Bloch
sphere picture in the sense of axes, with a multiplier effect similar to that
of unital qubit channels in the case of positive multipliers.    The inadequacy 
of this picture in the case of negative multipliers arises already for qubits.  
However, it is obscured by the unitary equivalence of maps composed 
with conjugation by a Pauli matrix $\sigma_k$.    For channels constant on
axes, this simple map is replaced by $\Psi_J^{\rm X}$, which is the 
 {\em average} of conjugations
with powers of the axis generators $W_J^\ell$, and
the picture for negative multipliers breaks down completely.

\bigskip


\appendix

\section{Convex combinations of unitary conjugations}  \label{app:pre}

\subsection{Orthogonal  bases of unitary operators}  \label{sect:OBU}

An {\em orthogonal basis of unitaries}  (OBU) for $M_d$ is a set of $d^2$ unitary
matrices $\{ V_0, V_1, \ldots V_{d^2 -1} \}$ with $V_0 = I$ satisfying
$\tr V_s^\dag V_t = d \, \delta_{st}$.    Since $M_d$ becomes a Hilbert
space when equipped with the inner product $\bra A, B \ket = \tr A^\dag B$,
one can expand an element of $M_d$ in this basis.   In particular,
any density matrix, $\rho$, which is a positive semi-definite operator
with $\tr \rho = 1$ can be written in the form
\be   \label{rhorep}
   \rho ~ = ~ \td \sum_{s = 0}^{d^2 -1}  v_s \, V_s ~ = ~ \td  \big[ I + \sum_{s > 0} v_s \, V_s \big]
\ee
with $v_s = \tr  V_s^\dag \rho$.   It follows easily \cite{FH,KMNR} that
\be   \label{rho.els}
    |v_s| \leq 1  \quad \text{and} \quad  \sum_{s > 0} |v_s|^2 \leq d-1
\ee
with equality in the latter if and only if $\rho$ is pure.   Although \eqref{rhorep}
can be regarded as a generalization of the Bloch sphere representation to
$d > 2$, the conditions in \eqref{rho.els} are necessary but not 
sufficient for an expression of the form  \eqref{rhorep} to define a positive
semi-definite operator.

Let $\{ | \psi_n \ket \}$ be an orthonormal basis for ${\bf C}^d$.    Then the
span of $\{ \proj{\psi_n } \}$  is a $d$-dimensional subspace of the 
$d^2$-dimensional space $M_d$. 
Now suppose that   $ \{ V_s \}$ is an
an OBU for $M_d$ such that  $\{ V_0, V_1, \ldots ,V_{d-1} \}$ span the same
subspace.   Then  the projections $\gamma_n = \proj{\psi_n } $ and the operators
 $  \tfrac{1}{\sqrt{d}} V_s $ with $ s =  0,1 \ldots d \mm 1$ give two orthonormal bases for  this
subspace.    Hence they are related by a unitary transformation, i.e.,
\be  \label{mutcom}
\gamma_n  = \td \sum_{s = 0}^{d}  x_{ns}  V_s  \qquad \hbox{and} \qquad
 V_s = \sum_{n = 1}^d  \overline{x_{ns}} \gamma_n
 \ee with
$ \tfrac{1}{\sqrt{d}} (x_{ns})$ unitary.      
Since the $\gamma_n$ commute, so do the $V_s$.  In fact, the vectors $|\psi_n \ket$ are 
simultaneous eigenvectors of these  $V_s$ with \eqref{mutcom} the
spectral decomposition.  This (or the purity condition \eqref{rho.els}) implies
that $|x_{ns}| = 1$ for all $n,s$. 

We will consider two special cases of an OBU in detail:
those associated with the generalized Pauli matrices introduced in
Section~\ref{sect:paul}, and those  associated  with generators of
mutually unbiased bases (MUB)  introduced in
Section~\ref{sect:mub}.   In both cases, each matrix $V_s$ will be
labeled by a pair of indices, so that $s \sim {(j,k)}$ or $s \sim (J,j)$. Despite its two indicies, $v_{jk}$ gives coefficients in a basis and is best regarded
as a column vector after some ordering of the indices rather than as
a matrix.  

\subsection{Representations of linear operators on $M_d$} \label{sect:OBU2}

When a linear operator $\Phi: M_d \mapsto M_d$
is represented by the $d^2 \times d^2$ matrix   $ T^{\Phi} $  with elements
\be  \label{maprep}
    T_{st} = \td \tr V_s^\dag \Phi(V_t),
\ee
its action on $\rho$ corresponds to $v_s \mapsto  \sum_t  T_{st} v_t$.
When  $T_{st} = \delta_{st}  \phi_t $ is a diagonal matrix, the channel is
called  {\em diagonal} and its action on $\rho$ reduces to
$v_s \mapsto \phi_s v_s$, i.e., it acts like a multiplier on the vector
representing $\rho$.

  If the unitary requirement is temporarily dropped and
$V_s \sim |e_j \kb e_k|$ in the standard basis for ${\bf C}^d$, then
$T_{(i,k),(j,\ell)} = \tr  |e_i \kb e_k| V( |e_j \kb e_{\ell}|)$ has the same
entries as the Choi-Jamiolkowski state representative but a very
{\em different} ordering!    It is important that the pair $(i,k)$ labels
rows and $(j,\ell)$ columns in order to correctly describe the action
of $\Phi$ by matrix multiplication.  The conversion from this ordering
to Choi-Jamiolkowski state form is sometimes called the
``canonical shuffle'' \cite{Paul}.

We are primarily interested in maps of the form
\be  \label{map:conj}
     \Phi(\rho) =   \sum_s  a_s V_s \rho V_s^\dag
\ee
with $a_s \ge 0$ and $\sum_s a_s = 1$.   Then $\Phi$ is a
unital completely positive, trace-preserving (CPT) map or 
unital quantum channel and $T_{0s} = T_{s0} = \delta_{0s}$.
\begin{thm}   \label{thm:diag}
Let $\{ V_s \}$ be an OBU satisfying a 
commutation relation of the form  
\be   \label{commcond}
  V_s V_t V_s^\dag V_t^\dag = \xi_{st} I.
\ee
Then  $|\xi_{st}| = 1$ and a channel  of the form \eqref{map:conj} is 
diagonal with multiplier  $\phi_s = \sum_u \xi_{su} a_u$.
 \end{thm}
A channel of the form
\eqref{map:conj} can be represented by a diagonal matrix even 
when the commutation condition  does not hold. However, when
only one $a_s$ is non-zero, i.e., $\Phi(\rho) = V_u \rho V_u^{\dag}$ 
for some fixed $u$, the channel is diagonal if and only if 
 \eqref{commcond} holds.
 
The next result may seem obvious; however, if the $V_s$ are not mutually
orthogonal, one can have a map of the form  \eqref{map:conj} which is
CP even though some $a_J$ are negative. An example is the qubit channel
\be
\Phi(\rho) = V \rho V^\dag = \rho + \sigma_x \rho \sigma_x - V^\dag \rho V 
\ee
where $V = \frac{1}{\sqrt{2}}\begin{pmatrix}1 & i \cr i & 1 \end{pmatrix}$. 
 \begin{thm}   \label{thm:aCP}
Let $\{ V_s \}$  be an OBU and $\Phi$ a map of the form
\eqref{map:conj}.   Then $\Phi$ is CP if and only if all $a_J \geq 0$.
\end{thm}
\prf   The key point is that when $\Phi(\rho) = U \rho U^\dag$ with $U$
unitary, its CJ matrix is the projection  $ \proj{U}$, where we employ a
slight abuse of notation in which $|U \ket$ denotes the $d^2 \times 1$
vectors obtained by ``stacking'' the columns of $U$.   When $\Phi$ has
the form \eqref{map:conj}, its CJ matrix can be written as
\be\label{eqn:aCP}
    \Gamma_{\Phi} = \td \sum_s  a_s \proj{V_s}.
\ee
Moreover, when the $V_s$ are mutually orthogonal, the corresponding
$|V_s \ket$ are also orthogonal and therefore eigenvectors of the CJ matrix
with eigenvalue $a_s$.    Thus,  $\Gamma_{\Phi}$ is positive semi-definite 
if and only if all $a_s \geq 0$.   \qed

\subsection{Generalized Pauli matrices}  \label{sect:paul}

In $d$ dimensions one can define the generalized Pauli matrices  $X$ and $Z$  
by their action on a fixed orthonormal basis   ${\bf C}^d$.
\be\label{def:XZ}
    X |e_k \ket = |e_{k+1} \ket    \qquad \hbox{and}  \qquad   
     Z |e_k \ket = \omega^k  |e_k \ket 
\ee
with  $\omega =   e^{i 2 \pi/d} $ and addition mod $d$ in the subscript.   
They are unitary and satisfy the commutation  relation $  ZX = \omega XZ$.
Thus, the  set of generalized Pauli operators
${\cal P} = \{ X^j Z^k : j,k = 0,1, \ldots d \mm 1\}$ forms an orthogonal
unitary basis for $M_d$.    We are interested in channels which have the
form \eqref{map:conj} in this basis, i.e., for which
\be  \label{Pdiag1}
    \Phi(\rho) = \sum_{jk} a_{jk} X^j Z^k \rho \, (X^j Z^k)^\dag
    \ee
with $a_{jk} \geq 0$ and $\sum_{jk} a_{jk} = 1$.    In view of Theorem~\ref{thm:diag},
the matrix   representing  $\Phi$ is a diagonal matrix; however, the diagonal
elements will not in general be  real.

 It is evident that $Z$ has the same properties as one of the $W_J$ in Section~\ref{sect:W}.
 In addition, $X$ and many other members of ${\cal P} $ are unitarily
 equivalent to $Z$ and share these properties.   
  Whenever $W = X^j Z^k$ with either $j$ or $k$ relatively prime to $d$, then $W$ generates a
cyclic group  of order $d$.
We want to exploit this group structure to relabel the matrices $X^j Z^k$ and
associate them with ``axes''  whenever possible.   For this purpose we do not
need to distinguish between, e.g.,  $(X^j Z^k)^2$ and $X^{2j} Z^{2k}$ although
$(X^j Z^k)^2 =  \omega^{jk} X^{2j} Z^{2k}$    With this notion of equivalence, we 
find that if $W_1 = X^{j_1} Z^{k_1}$ and $W_2 = X^{j_2} Z^{k_2}$ 
with  $gcd(j_1, k_1,d) = gcd(j_2, k_2,d) = 1$, then they   generate cyclic
groups $ {\cal W} _1$ and  ${\cal W}_2$ which are either equal or have no
common element other than $I$.

 Thus, when $d$ is prime,  the  set of generalized Pauli operators
${\cal P}$   can be partitioned
into the identity $I$ and $d+1$ disjoint sets of the form
$\{ W^j : j = 1, 2 \ldots d \mm 1\}$.  Let $W_K,  ~ K = 1, 2 \ldots d \pp 1$
denote some fixed choice of generators, and note that
$W_L  W_K =   \omega^t \, W_K  W_L$ where   $\omega = e^{2 \pi i/d} $
and $t$ is an integer which  depends on $L$ and $K$.   (One specific choice, used in
Appendix~\ref{app:CJ}, is  $W_J = X Z^J$ for $J = 1,2 \ldots d$ and $W_{d+1} = Z$.)
  The eigenvectors $\proj{\psi_n^J}$  of these $W_J$ form a set of $d \pp 1$ MUB.
  In view of \eqref{ax.state}, the  eigenstates of $W_J$ can be regarded as generalizations
of the qubit states $\half[I \pm \sigma_j]$ at the ends of the three  axes of the
Bloch sphere.   Thus, it is natural to call them  {\em axis states}.

When $d = d_1 d_2$ is not prime, then  $M_d \simeq M_{d_1} \ot M_{d_2}$ and
one can form another  OBU from tensor products of  generalized Pauli matrices 
in dimensions $d_1$ and $d_2$.     However, only when $d = p^m$ is a prime 
power is it known that one can make a similar division into MUB.
 
One might ask why we did not consider maps of the form \eqref{Pdiag1}
 with $a_s \sim a_{jk}$ independent of $k$:
 \be
      \Phi(\rho) = \td \sum_j a_j  \ \sum_k X^j Z^k \rho Z^{-k} X^{-j} = 
         \sum_j a_j X^j \Psi^{\rm QC}(\rho) X^{-j},
\ee
  We see that such a channel is a convex combination
of EB channels and, hence, also EB.   Therefore, this choice would not yield
  a particularly interesting new class of channels.

\subsection{Mutually unbiased bases}  \label{sect:mub}

A pair of orthonormal bases $\{ \vt_n \}$ and $\{ \psi_n \}$  is called 
mutually unbiased  if $\bra \vt_m, \psi_n  \ket = \td $.
A set of mutually unbiased bases (MUB) for ${\bf C}^d$ is a collection
of orthonormal bases $\{ |\psi_n^J \ket \},  ~ J = 1, 2, \ldots \kappa$ 
which are pairwise mutually unbiased as in \eqref{mubdef}.   
Our treatment of MUB is based on an  association with generators
$W_J$ as in \eqref{Wgen}.
\begin{thm}  \label{thm:MUBeqW}
A collection of orthonormal bases $\{ | \psi_k^J \ket \}$ is pairwise
mutually unbiased if and only if the associated generators satisfy
the orthogonality
condition    $ \tr W_J^{d-m} W_L^n = d \, \delta_{JL} \delta_{mn}$.
\end{thm}
\prf  One implication was shown in \eqref{Worth}.  The other follows
immediately from
\be     \label{Pcheck}
     |\bra  \psi_m^J,  \psi_n^L \ket |^2 & = &
        \tr \big( \proj{\psi_m^J} \big) \big(  \proj{\psi_n^L} \big)   \nn \\
   &  =  & \tfrac{1}{d^2} \sum_{j = 0}^{d-1}  \sum_{k = 0}^{d-1}  \ovb{ \omega}^{mj + kn} \, 
        \tr W^j_J W_L^k =  \tfrac{1}{d^2}  \tr I = \td.   \qquad
\ee
since  $\tr W^j_J W_L^k = \delta_{j0} \, \delta_{k0}$ when $J \neq L$.  \qed

As observed after \eqref{Worth}, one can have at most $d \pp 1$ MUB
for ${\bf C}^d$.   It follows immediately from Theorem~\ref{thm:MUBeqW} that
the existence of a maximal set set of $d \pp 1$ MUB  is equivalent
the existence of $d \pp 1$ unitary $W_J$ whose powers generate an OBU.
Moreover, this is equivalent to the existence of $d \pp 1$  
mutually orthogonal unitary $W_J$  with non-degenerate eigenvalues  
$\omega^k $ with $\omega = e^{2 \pi i/d}$.

The question of whether or not a maximal set of MUB
exist when $d$ is a composite of different primes is a difficult open
problem.  However, it is known that $d + 1$ MUB exist when $d = p^m$
is a prime power \cite{BBRV,KlappR1,WF}.      One method of constructing MUB 
is based on partitioning tensor products of Pauli matrices \cite{BBRV}.   
(See also \cite{HCM}.)
  \begin{thm}  \label{nice}
 When $d = p^m$ is a prime power, one can decompose the OBU
 formed by taking tensor products of generalized  Pauli matrices into
the identity and $d+1$ disjoint subsets of $d-1$ elements which we denote
$U_{Lj} $ with $  L =1,2 \ldots , d \pp 1 $ and $j = 1,2, \ldots d \mm 1$.   Moreover,
for each $L$,  $\{ I, U_{Lj}:j = 1, 2 , \ldots d \mm 1 \}$ forms a maximal Abelian 
subgroup of the Pauli group  and the
 simultaneous eigenvectors of the $U_{Lj}$ generate the MUB $\{ |\psi^L_n \ket \}$.
 \end{thm}
  
In this setting, the Abelian subgroups which define the MUB are not cyclic.
Although one can still use the MUB to define generators $W_J$, they need
not be equivalent to generalized Pauli matrices.
When $d$ is prime, the $W_J$ can be chosen to be generalized Pauli matrices
and, hence, satisfy a commutation relation.     
 \begin{qst}  \label{qst:cc}
Can the  generators $W_J$ for a fixed maximal set of MUB always be chosen 
so that they
satisfy  $W_J W_L = \xi_{JL} W_L W_J$ for some complex numbers
$\xi_{JL}$ with $|\xi_{JL}| = 1$?    
\end{qst}
Limited testing when $d = 4$ suggests that the answer is negative.
However, $W_J$ is not unique; it depends on the ordering of the basis.

Even when $d$ is not a prime power, one can find at least three
cyclic subgroups ${\cal W}_L$ with ${\cal W}_K \cap {\cal W}_L = I$ when
$K \neq L$.  One can choose as generators  $X, Z, XZ$, and define a set 
of three associated MUB.
\begin{qst}
 When $3 < \kappa < d \pp 1$, one can always extend $\{ W_L^j \} $ to an
 orthogonal basis for $M_d$.   Can this  be done so that the additional
 elements are also unitary?    
 
 When $\kappa = 3$, the generators $W_L$ can be chosen to be 
 generalized Pauli matrices, as in Section~\ref{sect:paul}. 
 Can generators of an MUB always be chosen to be either  
 generalized Pauli matrices
 or tensor products of  generalized Pauli matrices?  
 If they are always one or the other,  the answer to the previous question is positive.
  \end{qst}

\subsection{Channels based on MUB}  \label{app:MUBchan}

Channels constant on axes are special cases of channels of the form \eqref{map:conj}
with $V_s \sim  W_L^j$ and $a_s \simeq a_{Lj} = \tfrac{1}{d-1} a_L$ with
$a_L$ as in  \eqref{map:conjW}.  One could also consider $V_s \sim U_{Lj} $
 as  defined in terms of generalized Pauli matrices in Theorem~\ref{nice}; however, a channel of the form  \eqref{map:conj}
 with  $V_s \sim  W_L^j$ need not have this form with $V_s \sim U_{Lj} $.
If the coefficients depend on $j$ as well as $L$, the conversion could lead to
 cross-terms of the form $U_{Lj}  \rho U_{Lk} $ with $j \neq k$.   In fact,
 a channel of the form  \eqref{map:conj} with  $V_s \sim U_{Lj} $ is always diagonal,
 but one with $V_s \sim  W_L^j$ need not be.    However, for channels constant on
 axes, both are diagonal.

  For each fixed $L$, 
   $\{ W_L^j  \}_{j = 1, 2 , \ldots d - 1} $  and $\{ U_{Lj}  \}_{j = 1, 2 , \ldots d - 1} $  
 span the same subspace of $M_d$,  and many of the relations in Section~\ref{sect:2}
 can be written using $U_{Lj}$.   In particular
 \be   \label{QCeq}
  \Psi_L^{\rm QC}(\rho) =   \td  \sum_j U_{Lj} \rho \, U_{Lj}^\dag
      \ee
      and
  \be
         \proj{\psi_n^L} = \td \Big[ I +   \sum_{j = 1}^{d-1}   u_{Lj}  \,  U_{Lj} \Big].
 \ee
with $|u_{Lj} | = 1 $ and $\sum_j  | u_{Lj} |^2 = d-1$. 
We can also rewrite \eqref{rhorep} as
as
\be   \label{dbloch2}
      \rho     ~ = ~    \td \Big[ I + \sum_{J = 1}^{d + 1} \sum_{j = 1}^{d-1} u_{J j} U_{Jj} \Big] 
\ee
with $u_{J j} = \tr U_{J j}^\dag \rho$.

We are primarily interested in channels of the form \eqref{axcdef} when a full
set of $d \pp 1$ MUB exist.   However, even when only $\kappa < d \pp 1$ MUB exist, one can
generalize \eqref{axcdef} to 
 \be   \label{gen}
  \Phi = s \id + \sum_{L = 1}^{\kappa}   t_L \Psi_{L}^{\rm QC}+ u {\cal N}   .
 \ee
with the CPT conditions given by $s+ \sum_L t_L + u = 1$,    
$ s+  \td \sum_L t_L + \tfrac{1}{d^2} u \geq 0$,
and $    \td t_L  + \tfrac{1}{d^2} u \geq 0$.

When $d \pp 1$ MUB exist,  the completely noisy channel 
$\cN : \rho \mapsto (\tr \rho) \td I$ satisfies
\be  \label{noisy}
   \cN = \td \sum_L \Psi_{L}^{\rm QC} - \td \id, 
\ee
which allows one to reduce \eqref{gen} to \eqref{axcdef} by letting
$s \raw s - \td u$ and  $t_L = t_L + \td u$; in both forms
one has $\lambda_L = \wtd{s} + \wtd{t}_L$.    Even when 
$\kappa < d \pp 1$, one can associate a multiplier with the channel
\eqref{gen} by completing the orthogonal basis $W_J^k$. In this case:
 \be \phi_m = \begin{cases}
         s + t_L &  m \sim (L,j) \\
         s  & \hbox{otherwise}  \end{cases} \ee

 \subsection{Conjugations on a single axis} \label{app:sing_conj}

 We denote conjugation with a single unitary
 matrix  $U$ by $\Gamma_U$ so that  $\Gamma_U(\rho) \equiv U \rho U^\dag$.
When $U = U_{Jj}$ is an element of the OBU $\{ U_{Lk} \}$,
the channel $\Gamma_{U_{Jj}}$   is 
 diagonal in this basis with multiplier $\phi_s \sim \phi_{Lk} $ satisfying
 $|\phi_{Lk}| = 1$ and $\phi_{Jk} = 1$ for all $k$.    However, the map
 $\Gamma_{W_J^j}(\rho) = W_J^j \rho W_J^{-j}$ is {\em not} diagonal unless
 the commutation condition  \eqref{commcond}  holds as in Question~\ref{qst:cc}.
 When $d$ is prime, $U_{Jj} = W_J^j $ and we can say a bit more.
\be
  \Gamma_{U_{Jj}}(\proj{\psi^L_m})  =  \Gamma_{W_J^j}(\proj{\psi^L_m}) =
         \proj{\psi^L_n} 
   \ee 
where $n$ is a function of $L$ and $m$.  Thus,  $\Gamma_{W_J^j}$ 
permutes axis states when $d$ is prime.

It is useful to consider the special case of \eqref{map:conj} in which the unitary
conjugations involve only a single axis $J$.  The channels
\be   \label{Mdef}
      M_J(\rho)  = \sum_j c_j \Gamma_{U_{Jj}}(\rho) = \sum_j c_j U_{Jj} \rho U_{Jj}^\dag
\ee 
is  diagonal with multiplier satisfying $\phi_{Jj} = 1$ and
 $|\phi_{Lk}| \leq 1$ for $L \neq J$.    If $\Psi$ is a channel constant on axes, then a channel of the form
 $     \Phi =  \Psi \circ M_J $ still has a constant multiplier $\lambda_J$
 on the axis $J$ but has multipliers   $|\phi_{Lk}| \leq \lambda_L$ on
 on the other axes.   A channel of the form
 $    \Phi = x M_J + (1-x) \Psi $ , with  $0 \leq x \leq 1$
 has a constant multiplier $x + (1-x) \lambda_J$
 on the axis $J$, but has multipliers   $|\phi_{Lk}| \leq x + (1-x)\lambda_L$ on
 on the other axes.   Relaxing the requirement that the coefficients $a_{Lj}$
 are constant on one axis $J$ yields channels whose multipliers are 
 constant only on that axis.

\subsection{EB conditon on the $L_1$ norm of a channel}   \label{app:ccn}

We give a simple proof of the so-called ``computable cross-norm''
condition for separability.   This says that a bipartite density matrix
$\Gamma_{\Phi}$ is separable if and only if  $\sum_j \mu_j \leq 1$
when $\mu_j$ are the singular values after the canonical reshuffling  
of the elements so that  $\Gamma_{\Phi}$ is the CJ matrix of a CP map.
The conventions that $\tr \Gamma_{\Phi} = 1$, and $\Phi_{\Gamma}$
 satisfies the trace-preserving
condition $\tr \Phi_{\Gamma}(\rho) = \tr \rho$ are not consistent 
unless the reshuffling is accompanied by multiplication
by $d$.    Thus, theorem below gives an upper bound of $d$
rather than $1$.
\begin{thm}  \label{thm:ccn}
Let $\Phi$ be any EB channel.  Then $\norm{\Phi}_1 \leq d$.
\end{thm} 
\prf   It was shown in \cite{HSR} that a channel is EB if and only if it can be written in the
form $\Phi(\rho) = \sum_k R_k \tr \rho E_k$ where each $R_k$ is a density matrix
and $\{ E_k \}$ forms a POVM, i.e., each $E_k \geq 0$ and $\sum_k E_k = I$.
Then, with respect to the Hilbert-Schmidt inner product, $\Phi$ can be written as
\be
     \Phi = \sum_k | R_k \kb E_k |.
\ee
with the columns of $R_k$ and $E_k$ ``stacked'' as in (\ref{eqn:aCP}). 

Any positive semi-definite matrix satisfies $\tr E_k^2 \leq  (\tr E_k)^2$. Therefore,
it follows from the triangle inequality that
\be  \label{eq:ccn}
      \norm{\Phi}_1 & \leq & \sum_k \norm{ \, | R_k \kb E_k | \, }_1 
          =    \sum_k   ( \tr E_k^2)^{1/2}  ~(\tr R_k^2 \big)^{1/2}    \nn \\
          & \leq &   \sum_k   \tr E_k   
          = \tr I = d.   \qquad \qed
\ee

An immediate corollary applies to diagonal channels with $\ds\Phi\big(\sum_s a_sV_s\big) = \sum_s \phi_s a_sV_s$.
 \begin{thm}   \label{thm:EBT.pauli}
 Let $\Phi$ be a  channel which is diagonal in an OBU.
 If $\Phi$ is EB, then
 $\sum_s  |\phi_s| \leq d$.
 \end{thm}
 
 \section{Some multiplicativity proofs}  \label{app:mult}
 \subsection{Convex Combinations of Channels}  \label{app:elem}
The following elementary lemma is needed in Section~\ref{sect:posmult}
\begin{lemma}  \label{lemma:elem}
Let  $\Lambda_j$ be any set of channels for which $\nu_p(\Lambda_j) 
\leq B$ for all $j$,
and let $\Lambda $
be a convex combination of the $\Lambda_j$.  Then $\nu_p(\Lambda) \leq 
B$.  Moreover, if $\norm{\Lambda(\rho)}_p = B$ for some $\rho$, then 
$\nu_p(\Lambda) = B$.
\end{lemma}

As an illustration, we consider an application to channels with one
symmetry axis.  \begin{thm}  \label{thm:star}
Let  $\Phi(a_*,b_*) $be a channel of the form \eqref{planar} with $b_* 
> 0$
and $a_* < 0$.

i)  If $\nu_p\big[\Phi(a_*,b_*)\big] = \nu_p\big[\Phi(0,b_*)\big]$, 
then
$\nu_p\big[\Phi(a,b_*)\big] = \nu_p\big[\Phi(0,b_*)\big]$ for all $a 
\in (a_*,0)$.

ii)   If  (a) holds and, \eqref{mult} holds with  $\Phi = 
\Phi(a_*,b_*)$, then it also
holds for  $\Phi = (a,b_*)$ with $a \in (a_*,0)$.
\end{thm}
\prf  The proof follows immediately from  Lemma~\ref{lemma:elem} with
$B =  \nu_p\big[\Phi(0,b_*)\big]$ in part (a) and
$B =  \nu_p\big[\Phi(0,b_*)\big] \nu_p(\Omega)$ in part (b).
 
 \subsection{Conjecture \ref{conj:axisachieve} when $p=2$ and $p=\infty$}
 
 Conjecture \ref{conj:axisachieve} posits that the maximal output $p$-norm of a channel constant on axes is achieved on an axis state. We show this to be true in the special cases $p=2$ and $p=\infty$.
 
   \begin{thm}\label{lemma:infty}
     Let $ \Psi$ be a channel constant on axes with $| \lambda_L | \le |\lambda_{L^*}|$ for all $L$. Then 
     \be
     \nu_2(\Psi) &=& \norm{\Psi(\proj{\psi_0^{L^*}})}_2 \label{eqn:ConjP=2}\\
       \nu_\infty(\Psi) &=& \norm{\Psi(\proj{\psi_0^{L^*}})}_\infty \ee
     \end{thm}
     
 In the case $p=2$, this is uses the fact that the axis state saturates the inequality  \eqref{FKNup}  in Section \ref{sect:p2mult}. When the multipliers are all non-negative, we could prove the $p = \infty$ case using Theorem 15 from \cite{KMNR}. However, a more general proof using singular values works for all axis channels and, indeed, it seems likely that one could generalize it to all $p>2$. We present the proof for channels with $| \lambda_{L^*} | \ge \frac{1}{d}$; a similar argument works when $| \lambda_{L^*} | <  \tfrac{1}{d-1}(1-\lambda_{L^*})$.
 
 For a density matrix $\rho$, let $\{y_i\}$ be the singular values of $\Psi(\rho)$ with $y_0 = \norm{\Psi(\rho)}_\infty$. If $y_0 > | \lambda_{L^*} |$ then \be
 \norm{\Psi(\rho)}_2^2 &=& \sum_{i = 0}^{d-1} y_i^2 = y_0^2 + \sum_{i = 1}^{d-1} y_i^2 \\ &\ge& y_0^2 + \tfrac{1}{d-1}\Big(\sum_{i = 1}^{d-1} y_i\Big)^2 = y_0^2 + \tfrac{1}{d-1}(1 - y_0)^2 \\ &>& \lambda_{L^*}^2 + \tfrac{1}{d-1}(1 - \lambda_{L^*})^2 = \norm{\Psi(\proj{\psi_0^{L^*}})}_2^2 \ee
 which contradicts \eqref{eqn:ConjP=2}. 
 
 Therefore, $\norm{\Psi(\rho)}_\infty \le \norm{\Psi(\proj{\psi_0^{L^*}})}_\infty$ for all $\rho$. \qed
 
   \subsection{Critical points}  \label{app:crit}
 
 The following result  emerged from our study of channels constant on axes. We present it here in 
 full generality.
 \begin{lemma}  \label{lemma:crit}
Let $\Phi$ be a positivity-preserving  linear map on $M_d$ and
 $\{ | \psi_n \ket \}$  an orthonormal basis for ${\bf C}^d$   such that 
span$\{ \proj{\psi_n } \}_{n = 1,2 \ldots d}$ 
is an invariant subspace of both $\Phi$ and $\wh{\Phi}$.   Let $\gamma(t)$
be a differentiable one-parameter family of pure states with 
$\gamma(0) = \proj{\psi_m}$ for some $m$.   Then for all $p \geq 1$, the functions $ \nrm\Phi[\gamma(t)]\nrm_p$  and $S\big(\Phi[\gamma(t)]\big)$ each 
has a critical point at $t = 0$.
Moreover, if $\Phi_1$  and $\Phi_2$ are positivity-preserving  linear maps on $M_{d_1}$ 
and $M_{d_2}$ and each  satisfies the
same hypotheses, then the result holds for any differentiable  family
    $\gamma_{12}(t)$ of pure states on 
    ${\bf C}^{d_1 d_2}$ 
    with $\gamma_{12}(0) =  \proj{ \psi_{n}^1 \ot \psi_{m}^2 } $.
 \end{lemma}
 \prf   First, observe that since $\gamma(t)$ is a pure for all $t$, $\tr \gamma(t) =  \tr [\gamma(t)]^2 = 1$ is constant,
which implies $\tr \gamma'(t)  = 0$.    Then,
writing $\gamma(t) = \proj{\chi(t)}$, we see that 
 \bee
 \gamma'(t) &=& \vert \chi(t) \kb \chi'(t) \vert + \vert \chi'(t) \kb \chi(t) \vert.
 \eee
 Thus
 \be  \label{deriv}
    0 = \tr \gamma^\prime(0) = \bra  \chi'(0) ,  \psi_m \ket + \bra \psi_m , \chi'(0) \ket   = 2 \,  {\rm Re} \, \bra \psi_m , \chi'(0) \ket.
        \ee
Now let $f(t) = \tr \big( \Phi[\gamma(t)]\big)^p$ and observe that
 \be
    f^\prime(t) =  p \tr  \big( \Phi[\gamma(t)] \big)^{p-1} \Phi[\gamma^\prime(t)]   
       =  p  \tr \wh{\Phi} \big[ \big( \Phi[\gamma(t)]\big)^{p-1}\big] \gamma^\prime(t)  
       \ee
The invariance condition  on $\Phi$ implies that $\Phi[\gamma(0)] = \sum_n \mu_n \proj{ \psi_n}$
and that fact that it is positivity preserving implies that $\mu_n \geq 0$.
Then it follows from the orthogonality of the $\psi_n$ that 
$ \Phi[\gamma(0)] \big)^{p-1} =  \sum_n \mu_n^{p-1} \proj{ \psi_n}$.   Then using
the invariance of $\wh{\Phi}$,
we can find  $\wtd{\mu}_n$ such that
\be
      f^\prime(0) & = &   p  \tr \Big( \sum_n \wtd{\mu}_n \proj{ \psi_n} \Big) \gamma^\prime(t)  \nn \\
        & = &   p \tr \Big( \sum_n \wtd{\mu}_n \proj{ \psi_n} \Big) 
         \Big( \vert \psi_m \kb \chi'(0) \vert + \vert \chi'(0) \kb \psi_m \vert \Big)  \nn \\
         & = & 2 p \wtd{\mu}_m   \big( \bra \psi_m , \chi'(0) \ket +   \bra  \chi'(0) ,  \psi_m \ket \big) ~ = ~ 0
\ee
 by \eqref{deriv}.   A similar argument holds for $f(t) = S\big(\Phi[\gamma(t)]\big)$.  Note that $\Phi$
 positivity-preserving is needed only to ensure that $\big( \Phi[\gamma(t)] \big)^{p-1} $ and
 $\log \big( \Phi[\gamma(t)] \big)^{p-1} $ are well-defined and differentiable.    In the case of a tensor
 product, it suffices to observe that the channel $\Phi_1 \ot \Phi_2$ and the products $\{ | \psi_m^1 \ot \psi_n^2 \ket \}) $
 satisfy the hypotheses of the lemma for $d = d_1 d_2$. 
 \qed
 
 Note that because $M_d$ has dimension $d^2$ the hypothesis that the
 $d$-dimensional space
 span$\{ \proj{\psi_n } \}$ is invariant is far from trivial.    Special cases are given below.
 
 \begin{enumerate}
    \renewcommand{\labelenumi}{\theenumi}
    \renewcommand{\theenumi}{(\alph{enumi})}
 
\item  $\Phi$ is a channel constant on axes and$\{ | \psi_n^J \ket \}$ is one of the MUB.
  
\item $\Phi$ is a Pauli  diagonal channel,  and  $ \{ | \psi_n \ket \}$ are the  common eigenvectors
 of a commuting subset of $d$ generalized Pauli matrices.
When $d$ is not prime,  the  subgroup need not be cyclic.  
  For example, for $d = 4$, the lemma applies to the simultaneous eigenvectors of the set 
  $\{  I, X^2, Z^2, X^2 Z^2 \}$.
     
\item  Let $\Phi$ be a tensor product of unital qubit channels and
      $| \beta_n \ket = (I \ot \sigma_n) \vert \beta_0 \ket  $ be the four
      maximally entangled states (with $ | \beta_0 \ket    = |00 \ket + |11 \ket$).
      Then 
      \bee
      \hbox{span}\{\proj{\beta_n} \} =  \hbox{span} \{ I \ot I, \sigma_x \ot \sigma_x, 
      \sigma_y \ot \sigma_y, \sigma_z \ot \sigma_z \}
      \eee
and this is an invariant subspace of $\Phi$.   So for a product of unital qubit channels,
we have a critical point at the maximally entangled states.   Even when the inputs are not optimal,
this critical point can be a relative maximum; see the example in Fig. 6 of \cite{ERATO}.

\item  This lemma does not apply directly to the Werner-Holevo channel \cite{WH}
${\cal W}(\rho) = \frac{1}{d-1}\Big(I - \rho^T\Big) $ because
${\cal W}=\wh{{\cal W}}$ maps a basis to its complex conjugate, i.e., 
\be
{\cal W}\Big(\hbox{span}\{ \proj{\psi_n } \}\Big) = \hbox{span}\{ \proj{\ovb{\psi}_n}  \}
\ee
For any pure input,  
$\big( {\cal W}(\proj{\psi}) \big)^p = \tfrac{1}{(d -1)^p}{\cal W}(\proj{\psi}) $, and
$  \wh{{\cal W}}( \proj{\ovb{\psi}_n } ) \in $ span$\{ \proj{\psi_n } \}$ since $ \wh{{\cal W}} = {\cal W}$. 
Therefore, we have all that is needed for the proof of the statement,
\be
  \wh{{\cal W}}\big( {\cal W}(\proj{\psi_n}) \big)^p  \in \hbox{span}\{ \proj{\psi_n } \}
  \ee
  so that the conclusion still holds.

For a single use of $\Phi$, this affirms that any pure state is a critical point of the p-norm, which is clear since all pure state outputs have the same spectrum. For the product  ${\cal W} \ot {\cal W}$, this shows that any maximally entangled state is a critical point of both the output $p$-norm and entropy. 

\end{enumerate}

\subsection{Proof of Theorem~\ref{thm:FHN}}    \label{app:FHN}

Using the notation of Section~\ref{sect:OBU} with  $\{V_s \}$  the OBU,
 observe that   any $\rho_{12}$ can be written as
$\rho_{12} = \td \sum_s V_s \ot A_s $ with $A_s = \tr_1 (V_s^\dag \ot I) \rho_{12}$.
Then $\rho_2 = \td A_0$ and
\be
    (\Phi \ot \Omega)(\rho_{12}) = \td \sum_s \phi_s V_s \ot \Omega(A_s).
\ee
Then  defining
$|\phi_{\max}| = \sup_{s > 0} |\phi_s|$, one finds
\be
   \norm{   (\Phi \ot \Omega)(\rho_{12})}_2^2 & = &
       \tr   (\Phi \ot \Omega)(\rho_{12})^\dag   (\Phi \ot \Omega)(\rho_{12}) \nn  \\
              & = & \td \Big[ \tr |\Omega(A_0)|^2 + \sum_{s > 0} |\phi_s|^2  \,  \tr  |\Omega(A_s)|^2 \Big] \\
              & \leq &  \td \Big( \tr |\Omega(\rho_2)|^2  (1- |\phi_{\max}|^2  )
               + |\phi_{\max}|^2  \sum_s \tr  |\Omega(A_s)|^2 \Big)  \nn \\
              & \leq &    \td \Big( [\nu_2(\Omega)]^2(1- |\phi_{\max}|^2 )+
                          |\phi_{\max}|^2 d  \norm{(\id_1 \ot \Omega)(\rho_{12}) }_2^2 \Big)  \nn \\
              & \leq  &   \td \Big( [\nu_2(\Omega)]^2(1- |\phi_{\max}|^2 )+
                          |\phi_{\max}|^2 d  [\nu_2(\id \ot \Omega)]^2 \Big)  \nn \\    \label{FHNfin}
               & = & \td \big(    1 + (d-1)   |\phi_{\max}|^2 \big)      [\nu_2(\Omega)]^2  
\ee
where  we used the fact \cite{AHW} that
$\nu_2(\id \ot \Omega) = \nu_2(\Omega)$.   
When dim ${\mathcal H}_2 = 1$, \eqref{FKNup} follows.
Moreover, if the upper bound  \eqref{FKNup}  is attained, then 
\eqref{FHNfin} implies that
\be
 \norm{   (\Phi \ot \Omega)(\rho_{12})}_2^2 \leq    [\nu_2(\Phi)]^2   [\nu_2(\Omega)]^2
 \ee
and  this bound can always be attained by using a tensor product input.
    \qed

\section{Separability of some CJ matrices}  
\label{app:sep}
\subsection{Extreme points with one symmetry axis.}
\label{app:FHsep}
To prove Theorem \ref{thm:EB=PPT} in Section
\ref{sect:FHEB}, we need to establish that the points
$R$ and $Y$ in Figure~\ref{fig:FH.EB} correspond to
channels with separable CJ matrices. 

For the point $R = (-\td,\td)$, we will use a
construction due to P. Horodecki \cite{HorP2}
which extends an argument in \cite{HorP1}.   One can
verify that the CJ matrix
 \eqref{CJbeta} can be written as
\be
    \Gamma & = &  \tfrac{1}{d^2} \big( I  + \sum_{j
\neq k} |e_j \ot e_j \kb e_k \ot e_k | \big)    \\ \nn
            & = &  \tfrac{1}{d^2} \tfrac{1}{m^{d-1}}
\sum_{x_2 \ldots x_d}
              \proj{\phi_{x_2 \ldots x_d}  \ot
\ovb{\phi}_{x_2 \ldots x_d}}
\ee
where $m \geq 3$ is an integer,  $x_1 = 1$ and each of
the $d-1$ remaining
 $x_j$ is chosen from among the $m$-th roots of
unity $e^{2 \pi i n/m}$, the sum runs over all
possible choices of $x_2 \ldots x_d$, and
\bee
   | \phi_{x_2 \ldots x_{d}} \ket = 
\tfrac{1}{\sqrt{d}} \sum_{j = 1}^d x_j |e_j \ket
\eee
The point $Y = (\half,\tfrac{-1}{2(d-1)})$ corresponds
to the channel $\Psi_J^{\rm YEB}$, for which the CJ
matrix
 \eqref{CJbeta} can be written 
 \be  \label{CJeqY}
  \Gamma & = & \tfrac{1}{2 d(d-1)} \Big( \sum_{j \neq
k} \proj{e_j \ot e_k } +   \nn \\
     & ~ & \qquad   \qquad +  ~
      (d \mm 1) \sum_k \proj{e_k \ot e_k }  - \sum_{j
\neq k}  |e_j \ot e_j \kb e_k \ot e_k | \Big)  \nn \\
       & = &  \tfrac{1}{2 d(d-1)}  \sum_{j < k }    
\gamma_{jk}
 \ee
where $\gamma_{jk} $ is given by
\bee
 \proj{e_j \ot e_k } + \proj{e_k \ot e_j}   + ( | e_j
\ot e_j \ket -   | e_k \ot e_k \ket ) ( \bra e_j \ot
e_j |-   \bra e_k \ot e_k |) .
   \eee
 Each $\gamma_{jk} $ corresponds to a qubit density
matrix of the form
   \bee
      \pmx  1 & 0 & 0 & -1 \\ 0 & 1 & 0 & 0 \\ 0 & 0 &
1& 0 \\ -1 & 0 & 0 & 1 \emx
   \eee
which is separable because it satisfies the PPT
condition.    Thus, \eqref{CJeqY}
is a convex combination of separable matrices.

\subsection{State representatives for $d$  prime}
\label{app:CJ}
To further characterize the EB maps, we need more
information about
the CJ matrix.     We first consider only the case of
prime $d$, for
which the generators $W_J$ are generalized Pauli
matrices.
   There is no loss of generality
in assuming that  $W_J = X Z^J$ for $J = 1,2 \ldots d$
and that  $W_{d+1} = Z$.    For $J \ne d+1$:
\bee
     W_J^m  \big(|e_j \kb e_k| \big) W_J^{-m} =
\omega^{mJ(j-k)} |e_{j+m} \kb e_{k+m} |
\eee
so that
\be\label{PSI-X}
      \Psi^{\rm X}_J \big(|e_j \kb e_k| \big) =
          \tfrac{1}{d-1} \sum_{m \neq 0} 
\omega^{mJ(j-k)} |e_{j+m} \kb e_{k+m} |.
\ee
Note that this implies that for $J \neq d \pp 1$,
  the CJ matrix $\Gamma_{\Psi_{J}^{\rm X}}$ has  the coefficient
of $|e_j \ot e_j \kb e_k \ot e_k |$ equal to zero for
all $j,k$, which means that the maximally entangled
state $| \beta \ket$  is  in its kernel.   
The same is true for the CJ matrix of  the CP map
$\wh{\Phi}_{d + 1}  =\sum_{J \neq {d +1}} a_J \Psi^{\rm X}_J$ with $a_J > 0$.
When $d = 3$ we can write its CJ matrix explicitly as
\be  \label{CJspec}
\wh{ \Gamma}_{d+1} =  \tfrac{1}{6}  \pmx   0 & 0 & 0 \qquad  0 & 0 & 0 \qquad  0 & 0 & 0   \\
                0 & \alpha & 0 \qquad  0 & 0 & \ovb{z} \qquad  z & 0 & 0 \\
                  0 & 0 & \alpha \qquad  z & 0 & 0 \qquad  0 &  \ovb{z}  & 0  \\ ~ \\
                  0 & 0 & \ovb{z} \qquad  \alpha & 0 & 0 \qquad  0 & z & 0 \\
                  0 & 0 & 0 \qquad  0 & 0 & 0 \qquad  0 & 0 & 0 \\
                  0 & z & 0 \qquad  0 & 0 & \alpha \qquad   \ovb{z}  & 0 & 0  \\~~ \\
                  0 &  \ovb{z}  & 0 \qquad  0 & 0 & z \qquad  \alpha & 0 & 0 \\
                  0 & 0 & z \qquad   \ovb{z}  & 0 & 0 \qquad  0 & \alpha & 0 \\
                  0 & 0 & 0 \qquad  0 & 0 & 0 \qquad  0 & 0 & 0   \emx
\ee
where $z =   a_1 \, \omega +  a_2  \, \omega^2 + a_3$ and
 $\alpha = a_1 + a_2 +a_3$ which is $1$ when  $\wh{\Phi}_{d + 1}$ is   TP.

To obtain the general CJ matrix, observe
that we can use \eqref{conv} to write any channel constant on axes as
\be
    \Phi & =  & a_{00} \id + a_{d+1} \Psi_{d+1}^{\rm X} +
               \sum_{J = 1}^d a_J \Psi_J^{\rm X}    \nn \\
         & =  & \td \big[(d-1)s +1  -   t_{d+1}\big]
\id + t_{d+1}  \Phi^{\rm QC}_{d+1} +  \wh{\Phi}_ 
{d+1}    
    \ee
with $a_{00}, a_J, s, t_J$ related as following \eqref{map:conjW}.  
Then
    \be   \label{ebgam}
    \Gamma_{\Phi} = \td \big[(d-1)s +1  -   t_{d+1}\big]  \proj{\beta} 
+
       \tfrac{t_{d+1}}{d} \sum_j  \proj{e_j \ot e_j } +     \wh{\Gamma}_{d + 1} 
       \ee
We now give the nonzero elements of  $ \Gamma_{\Phi}$
with the conventions that indices with different letters are always  
unequal.
The first two come from the first two terms in \eqref{ebgam}
and the next two from $\wh{\Phi}_{d+1}  $ and \eqref{PSI-X}.
\bee \begin{array}{clll}
\hbox{Term:}  &\hbox{Coefficient:} & &  \\  ~~ & & & \\
|e_j \ot e_j \kb e_j \ot e_j |   &
   \tfrac{1}{d^2} \big[(d-1)s - (d-1) t_{d+1}   \big]  & = &
     \tfrac{1}{d^2} \big[1  + (d-1)\lambda_{d+1}\ \big]      \\
     |e_j \ot e_j \kb e_k \ot e_k |  & \tfrac{1}{d^2} \big[1 + ds -
\lambda_{d+1} \big]  & = &   \tfrac{1}{d^2}
\big[ \ds{\sum_{J=1}^d\lambda_{J} } \big] \\
\proj{e_j \ot e_k} & 
   \tfrac{1}{d(d-1)} \big( 1 - a_{00} - a_{d+1} \big) & = &
    \tfrac{1}{d^2} \big( 1 - \lambda_{d+1} \big) \\
|e_j \ot e_{j+m} \kb  e_k \ot e_{k+m} | &
\tfrac{1}{d^2} \ds{\sum_{J \ne d+1}  \omega^{mJ(j-k)} t_J }
& = &   \tfrac{1}{d^2} \ds{\sum_{J \ne d+1}  \omega^{mJ(j-k)} }
\lambda_J
\end{array} \eee

   We write the CJ matrix explicitly in the
case $d=3$:

 \be   \label{mat.eb2}
  \tfrac{1}{9}  \pmx   1\pp 2\lambda_4 & 0 & 0 &\quad 
0 &u & 0 &\quad  0 & 0 &u    \\
                0 & 1 \mm \lambda_4 & 0 &\quad 0 & 0 &
z &\quad \ovb{z} & 0 & 0 \\
                  0 & 0 & 1 \mm \lambda_4 &\quad
\ovb{z} & 0 & 0 &\quad 0 &  z  & 0  \\ ~ \\
                  0 & 0 & z &\quad  1 \mm \lambda_4 &
0 & 0 &\quad  0 & \ovb{z} & 0 \\
                u  & 0 & 0 &\quad  0 & 1\pp 2\lambda_4
& 0 &\quad  0 & 0 &u  \\
                  0 & \ovb{z} & 0 &\quad  0 & 0 & 1
\mm \lambda_4 &\quad   z  & 0 & 0  \\~~ \\
                  0 &  z  & 0 &\quad  0 & 0 & \ovb{z}
&\quad  1 \mm \lambda_4 & 0 & 0 \\
                  0 & 0 & \ovb{z} &\quad   z  & 0 & 0
&\quad  0 & 1 \mm \lambda_4 & 0 \\
                u   & 0 & 0 &\quad  0 &u  & 0 &\quad 
0 &   0  & 1\pp 2\lambda_4   \emx  \qquad
\ee

where $u = \lambda_1 + \lambda_2 + \lambda_3 $ and
$z  = \lambda_1 \, \omega + \lambda_2  \, \omega^2 +
\lambda_3$.
\subsection{Implications and Proofs for $d = 3$}  
\label{sect:d3}

\subsubsection*{Proof of Theorem~\ref{thm:PPT3}}
Applying the partial transpose to \eqref{mat.eb2}
gives a matrix which can be permuted to give three
similar $3 \times 3$
blocks so that the PPT condition is equivalent to
\be
 \tfrac{1}{9} \pmx
          1 + 2 \lambda_1 & \ovb{z} &  \ovb{z} \\   
z&  1- \lambda_1 & u \\
          z & u & 1 - \lambda_1 \emx
                \geq 0
\ee
\noindent Conjugating this with the unitary matrix
$\1rt2 \pmx \sqrt{2} & 0 &  0 \\
   0 & 1 & 1 \\ 0 & 1 & -1 \emx $ gives the equivalent
  condition
   \be
     \tfrac{1}{9} \pmx 1 + 2 \lambda_1 &
\sqrt{2}\ovb{z} & 0 \\
     \sqrt{2} z & 1 - \lambda_1 + u & 0 \\
     0 & 0 & 1 - \lambda_1 - u  \emx \geq 0.
   \ee  
This gives a pair of necessary and sufficient
conditions for PPT
  \bee
       1 - \lambda_1 - u  \geq 0  \qquad \hbox{and}
\qquad 
            (1 + 2 \lambda_1)(1-  \lambda_1) \geq 2
|z|^2
       \eee       
The first is equivalent to  $\sum_J \lambda_J     \leq
  1  $ which was   shown
in Theorem~\ref{thm:PPTd} to be necessary for all $d$;
the second is \eqref{ppt3}.   \qed

On the base tetrahedron, $\sum_J \lambda_J = - \half$
and the second condition
reduces to $\sum_J \lambda_J^2 \leq \tfrac{1}{4}$; 
this is the
equation of the inscribed sphere which just touches
the
faces of the base tetrahedron, as shown in
Figure~\ref{fig:base}

\subsubsection*{Extreme points}

Next consider a channel of the form $\wh{\Phi}_K =
\sum_{J \neq K}  a_J \Psi_J^\rmx$ which is a convex
combination of $d$ of the $\Psi_J^\rmx$. When $ d= 3$
and $K = d+1$,  
CJ matrix of $\wh{\Phi}_K $ is given by \eqref{CJspec}.
  It is easy to see that the PPT condition for
separability can never be satisfied unless
$z = 0$.   But this will happen if and only if all
$a_J = \td$.   This corresponds
to a channel $\Psi_J^{\rm XEB}$ in the center of one
of the faces of the ``base''
tetrahedron shown in Figure~\ref{fig:base}.   
Therefore,
$\Psi_J^{\rm XEB}$  is an extreme  point of the convex
hull of EB maps for $d = 3$.

The point $\Psi_J^{\rm YEB}$ lies on the line segment
  $t \Psi_J^{\rm XEB} + (1-t) \Psi_J^{\rm X}$, which
is the segment BE in Figure~\ref{fig:FH.EB}. 
  It follows from
  Theorem~\ref{thm:EB=PPT}  that these channels are EB
if and only if 
  $\half \leq t \leq 1$, i.e., that  $\Psi_J^{\rm
XEB}$ and $\Psi_J^{\rm YEB}$
  are the extreme points of the EB channels restricted
to this line segment. In the case  $d = 3$,
$\Psi_J^{\rm YEB} = \big[\tfrac{1}{4},-\tfrac{1}{4},
-\tfrac{1}{4},-\tfrac{1}{4}\big]$ lies on the boundary
of the sphere  $\sum_J \lambda_J^2 \leq \tfrac{1}{4}$
  which  encloses the set of PPT maps in the base
tetrahedron.  Since
  this set is strictly convex, $\Psi_J^{\rm YEB}$ must
be an extreme point
  of the subset of EB maps when $d = 3$.

Note that there are channels in the base tetrahedron
which
satisfy the CCN condition $\sum_J | \lambda_J | \leq
1$ but are not PPT.
For example, 
$\tfrac{1}{6}\Psi^{\rmx}_1 +  
\tfrac{2}{6}\Psi^{\rmx}_2 +  
\tfrac{3}{6}\Psi^{\rmx}_3$ gives
a channel with multiplier
$\big[ -\tfrac{1}{4}, 0,   + \tfrac{1}{4}, - \half 
\big]$.   This channel
has  $|\lambda_J|^2 = \tfrac{3}{8} > \tfrac{1}{4}$ and
lies on a face of the
 base tetrahedron with unequal  $a_J$.

\subsection{Some observations for $d > 3$:}\label{app:dprime}
The form of the CJ matrices  in
Appendix~\ref{app:CJ} generalizes to prime $d > 3$, 
in particular the pattern of non-zero elements in the matrix.   Indeed, the
only elements with non-zero coefficients must have 
the form
$|e_j \kb e_k | \ot |e_{j+m} \kb e_{k+m} |$.  When $j
= k$ or $m = 0$, the
coefficients were given explicitly after
\eqref{ebgam}.

Applying the PPT condition to a channel of the form
 $\wh{\Phi}_K = \sum_{J \neq K}  a_J\Psi_J^{\rm
X} $ with $ K = d \pp 1$
will yield $d \mm 1$ equations of the form
 $\sum_{J = 1}^d \omega^{n_J} a_J= 0$, with
 each of the $d$ roots of unity occurring exactly
once. We note
that setting all $a_J=  \td$ yields a solution
to these equations. 
Moreover, combining the  $d \mm 1$ PPT equations
  with the normalization condition $\sum_J a_J=
1$, gives
 $d$ equations for the $d$ numbers $a_J$.   The
coefficient matrix for
 the $a_J$ can be written in the form  $x_{jk} =
\omega^{jk}$.   This is a unitary
 matrix which implies that   $a_J= \td ~ \forall
J$ is the only solution.   Thus, as for
 $d = 3$, the only EB maps on the  ``faces'' of the
base are
  the $\Psi_J^{\rm XEB}$, which are thus true extreme
points of the
EB  subset of channels constant on axes.

We remark that the positivity of the submatrix of
$\Gamma_\Phi$
\bee
   \pmx 1 + (d \mm 1) \lambda_K  & u  & u & \ldots & u
\\
         u &  1 + (d \mm 1) \lambda_K  & u  & \ldots &
u  \\
         \vdots & & \ddots & & \vdots \\
         u & \ldots & & u &  1 + (d \mm 1) \lambda_K
\emx
   \eee
yields the inequalities \eqref{lamineq}.   In fact,
one eigenvalue is
\bee
     (d \mm 1)u + 1 + (d \mm 1) \lambda_K = 1 + (d \mm 1)\sum_J
\lambda_J
     \eee
     and the requirement that  this is $\geq 0$ gives
\eqref{lamineqb}.
    Considering a $2 \times 2$ submatrix gives
    $ ( 1 + (d \mm 1) \lambda_K)^2  \geq  u^2$ which
implies
    \eqref{lamineqa}. 
    
    \pagebreak

\noindent {\bf Acknowledgment:} It is a pleasure to 
acknowledge  helpful conversations  with many people,
including some with J. Emerson and D. Gottesman 
 about MUB;  with D. Bruss, J. Myrheim, M. Horodecki, P. Horodecki and
 others in Gdansk about separability criteria, and with
K. Audenaert  who did numerical tests of separability.
Much of this work was done when MBR was visiting the Perimeter
Institute in Waterloo and MN was at Kenyon College.

 \end{document}